\documentclass[aps,onecolumn,nopacs,nofootinbib,floatfix,superscriptaddress]{revtex4}
\usepackage{graphicx}
\usepackage{amsfonts}
\usepackage{amssymb}
\usepackage{amsbsy}
\usepackage{hyperref}
\usepackage{amsmath}
\usepackage{mathrsfs}
\usepackage{latexsym}
\usepackage{natbib}
\usepackage{bm}
\usepackage{subfigure} 
\usepackage{color}
\usepackage{wasysym}
\usepackage{mathbbol}
\usepackage{bigints}
\allowdisplaybreaks
\usepackage[normalem]{ulem}
\usepackage[dvipsnames]{xcolor}
\usepackage{multirow}

\usepackage{relsize}

\definecolor{napiergreen}{rgb}{0.16, 0.5, 0.0}

\begin{document}


\title{Equivalence in virtual transitions between uniformly accelerated and static atoms: from a bird’s eye}

\author{Pradeep Kumar Kumawat}
\email{pradeep.kumawat@iitg.ac.in}
\affiliation{Department of Physics, Indian Institute of Technology Guwahati, Guwahati 781039, Assam, India}

\author{Subhajit Barman}
\email{subhajit.barman@physics.iitm.ac.in}
\affiliation{Centre for Strings, Gravitation and Cosmology, Department of Physics, Indian Institute of Technology Madras, Chennai 600036, India}

\author{Bibhas Ranjan Majhi}
\email{bibhas.majhi@iitg.ac.in}
\affiliation{Department of Physics, Indian Institute of Technology Guwahati, Guwahati 781039, Assam, India}

\begin{abstract}
\noindent We study the prospect of the equivalence principle at the quantum regime by investigating the transition probabilities of a two-level atomic detector in different scenarios. In particular, two specific set-ups are considered. ($i$) {\it Without a boundary}: In one scenario the atom is in uniform acceleration and interacting with Minkowski field modes. While in the other the atom is static and in interaction with Rindler field modes. ($ii$) {\it With a reflecting boundary}: In one scenario the atom is uniformly accelerated and the mirror is static, and in the other scenario the atom is static and the mirror is in uniform acceleration. In these cases, the atom interacts with the field modes, defined in the mirror's frame. For both the set-ups, the focus is on the excitation and de-excitation probabilities in $(1+1)$ and $(3+1)$ spacetime dimensions. Our observations affirm that in $(1+1)$ dimensions, for both set-ups the transition probabilities from different scenarios become the same when the atomic and the field frequencies are equal. In contrast, in $(3+1)$ dimensions this equivalence is not observed in general, inspiring us to look for a deeper physical interpretation. Our findings suggest that when the equivalence between different scenarios is concerned, the excitation to de-excitation ratio provides a more consistent measure even in $(3+1)$ dimensions. We discuss the physical interpretation and implications of our findings.

\end{abstract}

\date{\today}

\maketitle

\section{Introduction}\label{sec:Introduction}

The Unruh and the Hawking effects are two of the major predictions of quantum field theory in curved spacetime that have pioneered the modern understanding of semi-classical gravity (see \cite{Hawking:1975vcx} and \cite{Unruh:1976db} for the original works on these effects). In the Unruh effect, a uniformly accelerated observer sees the Minkowski vacuum to be populated by a thermal distribution of particles. At the same time, in the Hawking effect, an asymptotic observer in a black hole background sees the event horizon emitting particles with a thermal spectrum. Although the Unruh effect corresponds to a flat background, while the other one relates to a curved background, they are intricately related among themselves. The fascinating similarity between these two events has long been attributed to the Equivalence principle (EP) (see \cite{Paunkovic:2022flx}, for a detailed discussion on different versions of the equivalence principle). For instance, according to the weak equivalence principle (WEP) \cite{Paunkovic:2022flx} -- the equation of motion of a particle is independent of the spacetime curvature and thus is given by geodesics. The implications of this WEP are evident when a static observer in a black hole spacetime is deemed to be similar to an accelerated observer in the flat spacetime, which also results in similarities \cite{Svidzinsky:2018jkp, Barman:2024dql} in their semi-classical characteristics. It is to be noted that there can also be dissimilarities \cite{Singleton:2011vh} in the semi-classical characteristics between these observers, and the validity of classical equivalence may not always imply quantum equivalence \cite{Zych:2015fka, Das:2023cfu}.

On the other hand, even in a flat background itself, there can be different equivalent scenarios when quantum particle creation is concerned. For instance, the spectrum when a uniformly accelerated observer sees the Minkowski vacuum is deemed similar to the spectrum when a static observer sees the Rindler vacuum \cite{Takagi:1986kn, Barman:2024dql}. In a similar direction, in $(1+1)$ dimensions it is also observed that the atomic excitation in a uniformly accelerated atom in the presence of a static mirror is equivalent to the excitation in a static atom in the presence of an accelerated mirror when the atomic and the field frequencies are the same \cite{Svidzinsky:2018jkp}. In this case, with the presence of a mirror, the classical equivalence in the relative motions between the atom and the mirror from the two scenarios is apparent. Thus any equivalence between these two scenarios in the quantum domain can be attributed to this classical equivalence, also reminiscent of Einstein's equivalence principle. However, there remain some crucial aspects to be verified. For instance, all of these scenarios regarding the equivalence between atomic transition probabilities are obtained in $(1+1)$ dimensions{\footnote{There are studies \cite{Chatterjee:2021fue,Das:2022qpx,Sen:2022cdx,Sen:2022tru} in $(1+1)$ dimensions where the inequivalence has been observed, but in different situations and setup.}}. While a $(3+1)$ dimensional scenario is expected to retain some of these characteristics, it is not guaranteed to what degree those equivalences can be validated. This concern is also presented in our previous work \cite{Barman:2024dql}, and in this work, we aim to answer some of these questions. In particular, we will concern ourselves with the following set-ups.
\begin{itemize}
    \item[$(i)~$] \textbf{Without a boundary:} In this set-up, in one scenario the atom is in uniform acceleration and it scans the Minkowski vacuum. At the same time, in the other scenario, the atom is static and scans the Rindler vacuum.
    
    \item[$(ii)~$] \textbf{With a reflecting boundary:} In this set-up, the atom is in the presence of a reflecting mirror. In one scenario, the atom is uniformly accelerated and the mirror is static, and in the other scenario the atom is static and the mirror is in uniform acceleration. The atom is always scanning the vacuum, defined in the mirror's frame.
\end{itemize}
In particular, in this work, we study the atomic excitation probabilities for these different set-ups and scenarios corresponding to the $(1+1)$ and $(3+1)$ dimensional spacetimes. In most cases the $(1+1)$ dimensional results are already known and we will only recall the final results. In those scenarios, we will focus on deriving the $(3+1)$ dimensional atomic excitation probabilities. As we will see in $(3+1)$ dimensions, which is more viable from a practical point of view, the expressions for the atomic excitation probabilities corresponding to different scenarios of the same set-up do not match when the atomic and the field frequencies are equal. This motivated us to also investigate the de-excitation probabilities for these different scenarios. We see that these de-excitation probabilities too suggest the same outcomes as the excitation probabilities. Furthermore, we estimate the excitation-to-de-excitation ratios (EDR) in each scenario, which can be relevant from a practical point of view. A possible argument is elaborated in the main text of the manuscript. We observe that in terms of EDR even in $(3+1)$ dimensions different scenarios are equivalent, making it a more suitable quantity for investigations in this direction.

The manuscript is organized in the following manner. In Sec. \ref{sec:Coords} we introduce the coordinate transformations between a uniformly accelerated observer and Minkowski frame and recall the set-up to investigate the atomic transition probabilities while the atom interacts with a neutral photon field from the background. In the subsequent sections \ref{sec:AccAtom-E-WOutBound} and \ref{sec:AccAtom-DE-WOutBound} we respectively investigate the atomic excitation and de-excitation probabilities without the presence of a boundary. In these sections both the $(1+1)$ and $(3+1)$ dimensional scenarios are studied. In a similar manner, in Secs. \ref{sec:AccAtom-E-WBound} and \ref{sec:AccAtom-DE-WBound} we investigate the atomic excitation and de-excitation probabilities respectively when the atom is in the presence of a reflecting mirror in both $(1+1)$ and $(3+1)$ dimensions. In Sec. \ref{sec:Ex-Deex-ratio} we estimate the excitation to de-excitation ratios for each of the different scenarios, e.g., when the atom is accelerated or the mirror is accelerated (similar situation without a boundary is obtained when the expectation taken in the Rindler vacuum), and for different spacetime dimensions. We present our main observations and discuss the implication of our findings in Sec. \ref{sec:discussion}.

\section{Rindler coordinate transformations and the set-up}\label{sec:Coords}
In this section, we recall the coordinate transformations appropriate for a uniformly accelerated observer in a Minkowski spacetime. In particular, we consider the observer to be accelerated along the positive or negative $z$-direction and thus all other spatial coordinates remain unchanged. Depending on whether the observer is accelerated along the positive or the negative spatial direction, its motion remains confined to the regions of a specific Rindler wedge (see \cite{Crispino:2007eb}) - the right Rindler wedge (RRW) or the left Rindler wedge (LRW). In this section, we shall also elucidate the set-up for obtaining the transition probability of a two-level atom that interacts with a background scalar photon field.

\subsection{Coordinate transformations}

Let us first recall the coordinate transformations relating the Minkowski $(t,\,z)$ and the right Rindler wedge coordinates $(\tau,\,\xi)$. These coordinate transformations are the same for both the $(1+1)$ and $(3+1)$ dimensional scenario with the other spatial coordinates $(x,\,y)$ remaining unchanged in $(3+1)$ dimensions. In particular, for an observer in uniform acceleration along the positive $z$ direction, the coordinate transformations \cite{Crispino:2007eb, Olson:2010jy, Quach:2021vzo, Barman:2024dql} relating the Minkowski coordinates $(t,\,z)$ and the Rindler ones $(\tau,\,\xi)$ are given by
\begin{eqnarray}\label{eq:CT-R}
    t &=& \frac{c}{a}\,e^{a\,\xi/c^2} \sinh{(a\,\tau/c)}~;~~z = \frac{c^2}{a}\,e^{a\,\xi/c^2} \cosh{(a\,\tau/c)}~.
\end{eqnarray}
This specific observer is confined to RRW. 
At the same time, for an observer in uniform acceleration along the negative $z$ direction, it is confined to LRW. The coordinate transformations from Minkowski to LRW $(\bar{\tau},\,\bar{\xi})$ are
\begin{eqnarray}
    t &=& -\frac{c}{a}\,e^{a\,\bar{\xi}/c^2} \sinh{(a\,\bar{\tau}/c)}~;~~z = -\frac{c^2}{a}\,e^{a\,\bar{\xi}/c^2} \cosh{(a\,\bar{\tau}/c)}~.\label{eq:CT-L}
\end{eqnarray}
In both of the above two coordinate transformations to RRW and LRW the parameter $a$ corresponds to the proper acceleration of an observer sitting at the origin of the Rindler frame, and $c$ denotes the speed of light. In the subsequent analysis, we shall utilize these coordinate transformations to obtain the transition probabilities in different scenarios.

\subsection{Atomic response: the set-up}

In this section, we consider a two-level atom interacting with a neutral photon field from the background. The normalized mode function for this field is denoted by $\phi$. Our set-up is broadly comprised of these atoms and the field, and shall follow the formulation of \cite{Svidzinsky:2018jkp,Chakraborty:2019ltu} to understand the atomic transition probabilities. The interaction Hamiltonian $\hat{H}_{I}$ corresponding to the atom-photon interaction is given by
\begin{eqnarray}\label{eq:H-int}
    \hat{H}_{I}(\tau) &=& \hbar\,g\Big(\phi_{\nu}\big[t(\tau),z(\tau)\big]\,\hat{a}_{\nu}+h.c.\Big)\,\big(e^{-i\,\omega\tau}\,\hat{\sigma}+h.c.\big)~.
\end{eqnarray}
In the above expression, $\omega$ and $\nu$ are respectively the atomic and field frequencies, $\hbar$ is the Planck constant, and $g$ denotes the interaction strength. The operators $\hat{a}$ and $\hat{\sigma}$ respectively denote the field and atomic lowering operators. $\tau$ denotes the clock time of an observer co-moving with the atom. We consider the atomic states $|\omega_{0}\rangle$ and $|\omega_{1}\rangle$ denote the ground and the excited states, and the field is initially prepared in the ground state $|0\rangle$. In particular, we shall recall the expression of the atomic excitation probability when the atom is prepared in its ground state, and also provide the expression for de-excitation probability if the atom is in its excited state. When the atom is in its ground state $|\omega_{0}\rangle$, its excitation probability with simultaneous emission of a photon of frequency $\nu$, which is due to $\hat{a}_{\nu}^{\dagger}\,\hat{\sigma}^{\dagger}$ in the interaction Hamiltonian, is
\begin{eqnarray}\label{eq:TP-exc-gen}
    \mathcal{P}_{\nu}^{ex}(\omega) &=& \frac{1}{\hbar^2} \bigg|\int d\tau\, \langle 1_{\nu},\omega_{1}|\hat{H}_{I}(\tau)|0,\omega_{0}\rangle\bigg|^2~= g^2\bigg|\int_{-\infty}^{\infty} d\tau\,\phi^{\star}_{\nu}\big[t(\tau),z(\tau)\big]\,e^{i\,\omega\,\tau}\bigg|^2~.
\end{eqnarray}
As previously mentioned this set-up and the expression for the above atomic excitation probability were initially provided in \cite{Svidzinsky:2018jkp}. At the same time, when the atom is in its excited state, its de-excitation with simultaneous emission of a photon of frequency $\nu$, which is due to the term $\hat{a}_{\nu}^{\dagger}\,\hat{\sigma}$ in the interaction Hamiltonian, is
\begin{eqnarray}\label{eq:TP-deex-gen}
    \mathcal{P}_{\nu}^{de}(\omega) &=& \frac{1}{\hbar^2} \bigg|\int d\tau\, \langle 1_{\nu},\omega_{0}|\hat{H}_{I}(\tau)|0,\omega_{1}\rangle\bigg|^2~= g^2\bigg|\int_{-\infty}^{\infty} d\tau\,\phi^{\star}_{\nu}\big[t(\tau),z(\tau)\big]\,e^{-i\,\omega\,\tau}\bigg|^2~.
\end{eqnarray}
In our forthcoming analysis, we shall utilize these expressions from Eqs. \eqref{eq:TP-exc-gen} and \eqref{eq:TP-deex-gen} to obtain the excitation and de-excitation probabilities of atoms in different trajectories corresponding to with and without the presence of a reflecting boundary.

Here we would like to mention the striking similarity between the integrals inside the above expressions and the expressions of the Bogoliubov-like coefficients as prescribed by Padmanabhan (Paddy), see Sec. $4$ of  \cite{Singh:2013dia}. One can notice that the integral of \eqref{eq:TP-exc-gen} is an exact complex conjugate of Paddy's $\beta$ coefficient. Whereas, the integral from Eq. \eqref{eq:TP-deex-gen} is the complex conjugate of Paddy's $\alpha$ coefficient. Therefore, the question of equivalence can also be discussed with these Bogoliubov-like coefficients.

\section{Atomic excitation without a boundary}\label{sec:AccAtom-E-WOutBound}
In this section, we estimate the atomic excitation probability in the absence of a boundary. First, we shall recall the results in $(1+1)$ dimensions, which are pretty straightforward. Second, we will estimate the excitation probabilities in $(3+1)$ dimensions, which as we will see, may contain interesting insights. We would also like to mention that in both the considered dimensions, we will estimate the transition probabilities considering: I. the atom is in uniform acceleration and the field vacuum is the Minkowski vacuum, II. the atom is static and the field vacuum is the Rindler vacuum. Therefore, in the second scenario, a static atom is expected to report excitation if it observes the Rindler vacuum. We will also check whether these two scenarios (I and II) are equivalent to each other.

\subsection{Excitation probability in $(1+1)$ dimensions}

In this section, we consider the $(1+1)$ dimensional scenario. Furthermore, first, we consider the situation when the atom is in uniform acceleration, i.e., the proper time of the atom is defined in a frame comoving with the atom which is in the Rindler frame. The field vacuum $|0\rangle$ is the Minkowski vacuum $|0_{M}\rangle$. 
In $(1+1)$ dimensions the scalar field mode solution is $\phi_{\nu} (t,z)=e^{- i\,\nu\, t+i\, k\, z}/\sqrt{4\pi\nu}$, see \cite{book:Birrell,Takagi:1986kn, Crispino:2007eb}. In the previous expression of field mode solution $\nu$ denotes the frequency of the mode and $k$ the wave vector, and they are related among themselves via the dispersion relation $|k|=\nu/c$.
We utilize this expression of the field mode in Eq. \eqref{eq:TP-exc-gen} to evaluate the excitation probability and get
\begin{eqnarray}\label{eq:Pex-Atom-Acc0}
    \mathcal{P}^{ex}_{\nu}(\omega) &=& \frac{g^2}{4\pi\,\nu}\bigg|\int_{-\infty}^{\infty} d\tau\,e^{- i\,\nu\, t+i\, k\, z}\,e^{i\,\omega\,\tau}\bigg|^2\nonumber\\
    &=& \frac{g^{2}c^2}{4\pi\,\nu\,a^2}\,\bigg|\int_{0}^{\infty} dx\,x^{-1- i\,\omega\,c/a}\,e^{-i\,\nu\,c\,x/a}\bigg|^2~.
\end{eqnarray}
To arrive at the previous expression, we have utilized the Minkowski to Rindler coordinate transformation of Eq. \eqref{eq:CT-R}, and the relation $t-z/c=-(c/a)\,e^{-a\tau/c}$. We have further considered the change of variables $x=e^{-a\tau/c}$. One can easily carry out the integration of the second line in Eq. \eqref{eq:Pex-Atom-Acc0}, which is obtained in terms of the Gamma function. Then the final form of the excitation probability from Eq. \eqref{eq:Pex-Atom-Acc0} can be expressed as
\begin{eqnarray}\label{eq:Pex-Atom-Acc}
    \mathcal{P}^{ex}_{\nu}(\omega) 
    &=& \frac{g^{2}c}{2\,a\,\nu\,\omega}\frac{1}{e^{2\pi\omega c/a}-1}~.
\end{eqnarray}

Second, we consider the situation when the atom is static in the Minkowski background but it responds to the Rindler vacuum fluctuations. In this scenario, the atomic proper time $\tau$ is described by the Minkowski time coordinate $t$, and the field vacuum $|0\rangle$ is the Rindler vacuum $|0_{R}\rangle$. 
In $(1+1)$ dimensions the scalar field mode solution in terms of the Rindler coordinates is $\phi_{\nu} (\tau,\xi)=e^{- i\,\nu\, \tau+i\, k\, \xi}/\sqrt{4\pi\nu}$, see \cite{book:Birrell}. Here also we have considered the frequency of the mode as $\nu$ and the wave vector as $k$, as it will help us compare the current scenario with the previous one. We utilize the expression of this field mode with \eqref{eq:TP-exc-gen} to estimate the atomic excitation probability, which in this case becomes
\begin{eqnarray}\label{eq:Pex-Atom-Stat0}
    \mathcal{P}^{ex}_{\nu}(\omega) &=& \frac{g^2}{4\pi\,\nu}\bigg|\int dt\,e^{- i\,\nu\, \tau +i\, k\, \xi}\,e^{i\,\omega\,t}\bigg|^2\nonumber\\
    &=& \frac{g^{2}}{4\pi\,\nu}\,\bigg|\int_{-\infty}^{0} dt\,\Big(-\frac{a\,t}{c}\Big)^{- i\,\nu\,c/a}\,e^{i\,\omega\,t}\bigg|^2~.
\end{eqnarray}
For an atom kept static in the Minkowski spacetime the proper time is the Minkowski time $t$, and that is why in the previous expression we have integrated over the time $t$. We have arrived at the second line of the previous expression utilizing the coordinate transformation of Eq. \eqref{eq:CT-R} and the relation $\tau-\xi/c=-(c/a)\ln{\big[a(-t+z/c)/c\big]}$. Furthermore, we considered that the atom is static at $z=0$. Then the previous relation becomes $\tau-\xi/c=-(c/a)\ln{[-a\,t/c]}$, and one can notice that the range of $t$ should be restricted to $t\in (-\infty, 0]$ to properly describe the interaction with Rindler frame. The integration inside the second line of the above expression \eqref{eq:Pex-Atom-Stat0} is obtained in terms of the Gamma function, and one can express the final result as
\begin{eqnarray}\label{eq:Pex-Atom-Stat}
    \mathcal{P}^{ex}_{\nu}(\omega)=\frac{g^{2} c}{2\,a\,\omega^2} \frac{1}{e^{2\pi\nu c/a}-1}~.
\end{eqnarray}
From the above two expressions of Eqs. \eqref{eq:Pex-Atom-Acc} and \eqref{eq:Pex-Atom-Stat} corresponding to an accelerated and a static atom respectively seeing the Minkowski and the Rindler vacuum, we observe that the excitation probabilities are described by Planckian distribution (modified by a multiplicative factor) with respect to the atomic $(\omega)$ and field $(\nu)$ frequencies, i.e., they are thermal with respect to the frequencies $\omega$ and $\nu$. However, it should be noted that in both cases the characteristic temperatures of these thermal distributions are proportional to the acceleration $a$.
These two excitation probabilities become the same when the atomic and the field frequency become the same, i.e., when $\omega=\nu$. This is well known, although probably not mentioned explicitly.

\subsection{Excitation probability in $(3+1)$ dimensions}
In this part of the section, we consider the $(3+1)$ dimensional scenario. The coordinate transformation to the Rindler frame remains identical to Eq. \eqref{eq:CT-R}, with the spatial coordinates $\vec{x}_{\perp}\equiv(x,\,y)$ remaining unchanged by this transformation. In $(3+1)$ dimensions we shall first consider the situation when the atom is in uniform acceleration and sees the Minkowski vacuum. Then we shall consider the situation when the atom is static and sees the Rindler vacuum.

\subsubsection{Uniformly accelerated atom}
In this part, we consider the atom to be in uniform acceleration along the positive $z$-direction. Furthermore, if we consider the atom to be stationed at the origin of the Rindler frame, i.e., at $\xi=0$, the coordinate transformation of Eq. \eqref{eq:CT-R} relating the Minkowski frame with the accelerated observer will become  
\begin{eqnarray}\label{eq:CT-R-2}
    t(\tau) = \frac{c}{a} \sinh{\left(a\,\tau/c\right)}~;~~z(\tau)= \frac{c^2}{a} \cosh{\left(a\,\tau/c\right)}~~.
\end{eqnarray}
In $(3+1)$ dimensional Minkowski spacetime the expression for the normalized field mode with frequency $\nu$ and wave vector $\vec{k}$, see Eq. (2.80) of \cite{Crispino:2007eb} and Eq. (2.11) of \cite{book:Birrell}, is
\begin{equation}\label{eq:Phi-3p1-WOB}
    \phi_{\nu}=\frac{1}{\sqrt{(2\pi)^3\,2\,\nu}}e^{-i\nu t+i\vec{k}_{\perp}.\vec{x}_{\perp}+i k_{z}z}~.
\end{equation}
It is to be noted that as we are dealing with a neutral photon field the wave vector and the frequency are related among themselves via the dispersion relation $|\vec{k}|=|k|=\nu/c$. We utilize the above expression of the mode function and coordinate transformation from Eqs. \eqref{eq:Phi-3p1-WOB} and \eqref{eq:CT-R-2} in Eq. \eqref{eq:TP-exc-gen} to obtain the excitation probability
\begin{eqnarray}\label{eq:3p1-atomic-atomAcc-4a}
        \mathcal{P}^{ex}_{\nu}(\omega) &=& \frac{g^2}{(2\pi)^3\,2\,\nu}\,\bigg|\int d\tau\,e^{i\nu\,t+i\omega\,\tau-i k_{z}z}\bigg|^2~\nonumber\\
        ~&=& \frac{g^2}{(2\pi)^3\,2\,\nu}\,\bigg| \int_{0}^{\infty}\frac{c}{a}\frac{d\lambda}{\lambda}  \lambda^{\frac{i\omega c}{a}}\exp\biggl\{ \frac{i\nu c}{2a}(1-\cos{\theta})-\frac{i\nu c}{2a}(1+\cos{\theta})\frac{1}{\lambda}\biggl\}\bigg|^2~\nonumber\\
        ~&=& \frac{g^2}{(2\pi)^3\,2\,\nu}\,\bigg| \frac{2c}{a}\,e^{-\frac{\pi\omega\,c}{2a}}\biggl(\frac{1+\cos{\theta}}{1-\cos{\theta}}\biggl)^{\frac{i\omega\,c}{2a}}\,\mathcal{K}_{\frac{i\omega\,c}{a}}\Big(\frac{\nu c}{a}\sin{\theta}\Big)\bigg|^2~.
\end{eqnarray}
In the second line of the previous equation, we have used the coordinate transformation \eqref{eq:CT-R-2}, expressed $k_z=|k|\cos\theta=(\nu/c)\,\cos\theta$, and considered a change of integration variables $e^{a\tau/c}=\lambda$. We should mention that in the last line of Eq. \eqref{eq:3p1-atomic-atomAcc-4a}, $\mathcal{K}_{\mu}(x)$ denotes the modified Bessel function of the second kind of order $\mu$. For the purpose of analytical analysis, we consider a large acceleration limit. For $x\ll 1$ one can series expand the above Bessel function, and this asymptotic expansion is given by (see \cite{gradshteyn2007} or Eq. (A10) of \cite{Crispino:2007eb})
\begin{eqnarray}\label{eq:BesselK-series-exp}
    \mathcal{K}_{\mu}(x) &\simeq& \frac{1}{2}\bigg[\Big(\frac{x}{2}\Big)^{\mu}\,\Gamma(-\mu)+\Big(\frac{x}{2}\Big)^{-\mu}\,\Gamma(\mu)\bigg]~.
\end{eqnarray}
For large acceleration $\nu\,c/a\ll 1$, we utilize the asymptotic expansion \eqref{eq:BesselK-series-exp} in Eq. \eqref{eq:3p1-atomic-atomAcc-4a} and obtain
\begin{eqnarray}\label{eq:3p1-atomic-atomAcc-4a-2}
        \mathcal{P}^{ex}_{\nu}(\omega) &\simeq&  \frac{g^{2}c}{2\pi^2\,\nu\, \omega\, a}\, \frac{\cos^{2}{\varphi_{ex}}}{e^{2\pi\omega c/a}-1} ~.
\end{eqnarray}
In the previous expression $\varphi_{ex}$ is given by $\varphi_{ex}=\mathrm{Arg}\big[\big\{c\nu\sin{\theta}/(2a)\big\}^{-i \omega c/a}\,\Gamma\big(i\omega c/a\big)\big]$, where the argument of a complex quantity $\kappa$ is defined as $\mathrm{Arg}\,[\kappa]=\tan^{-1}\left\{\mathrm{Im}(\kappa)/\mathrm{Re}(\kappa)\right\}$. For large acceleration we have $\nu\,c/a\ll 1$ and $\omega\,c/a\ll 1$ and we can further expand $\varphi_{ex}\approx \frac{\pi}{2}-\{{a/(\gamma\omega\,c)-\omega\,c\left( \gamma^{2} + \pi^{2}/6 \right)/(2\gamma\,a)}\}^{-1}$. Here $\gamma$ is the Euler–Mascheroni constant or the Euler's constant, which has the value $\gamma\approx0.5772$. Then the excitation probability becomes
\begin{eqnarray}\label{eq:3p1-atomic-atomAcc-4a-3}
    \mathcal{P}^{ex}_{\nu}(\omega)  &\simeq&  \frac{g^{2}c}{2\pi^2\,\nu \omega a}\, \frac{1}{e^{2\pi\omega c/a}-1} \,\sin^{2}{\left(\frac{1}{\frac{a}{\gamma\omega\,c}-\frac{1}{2}\left(\gamma^{2}+\frac{\pi^{2}}{6}\right)\frac{\omega\,c}{\gamma\,a}}\right)}~.
\end{eqnarray}

\subsubsection{A Static atom at $z=z_{0}$}

In this part, we consider a static atom at $z=z_{0}$ and check whether it can get excited while there is a simultaneous emission of a photon in the Rindler frame. In this regard, we consider the coordinate transformation Eq. \eqref{eq:CT-R}, and thus the line element will read as
\begin{equation}\label{eq:Rindler-LineEl}
    ds^{2} =e^{2a\xi/c^2}\Big[-c^2\,d\tau^2+d\xi^2\Big]+d\vec{x}^2_{\perp}~.
\end{equation}
This line element is utilized to obtain the field equation of motion $\partial_{\mu}(\sqrt{-g}\,g^{\mu\sigma} \partial_{\sigma}\phi)=0$ and then the field mode solution $\phi=\phi_{\nu,k_{\perp}}^{R}$, where $g^{\mu\sigma}$ and $g$ respectively denote the inverse and determinant of the metric tensor. The normalized field mode solution $\phi_{\nu,k_{\perp}}^{R}$, see \cite{Crispino:2007eb, Barman:2021oum}, in this scenario is expressed as
\begin{equation}\label{eq:Rind-FldMode-3p1}
    \phi^{R}_{\nu,k_{\perp}} =\sqrt{\frac{c\,\sinh{(\pi\nu \,c/a)}}{4\pi^4a}}\,\mathcal{K}_{i\,\nu c/a}\bigg(\frac{k_{\perp}e^{a\,\xi/c^2}}{a/c^2}\bigg)~e^{i\,\,\vec{k}_{\perp}.\vec{x}_{\perp}-i\,\,\nu\,\tau}~.
\end{equation}
We make use of the asymptotic expansion of the Bessel function from Eq. \eqref{eq:BesselK-series-exp} in large acceleration limit to obtain the field mode expansion. It is to be noted that after this asymptotic expansion, the field mode will have both the right-moving ($e^{i\,\nu\,\xi/c}$) and left-moving ($e^{-i\,\nu\,\xi/c}$) parts. The right moving parts correspond to incoming waves with respect to the mirror in RRW, and these parts will interact with the atom when $z>ct$. Thus they should be accompanied by a multiplicative factor of $\Theta(z-ct)$ when represented in the field mode, where $\Theta(x)$ denotes the Heaviside step function. Similarly, the left moving parts signify reflected waves from the mirror, and they interact with the atom when $z>-ct$. Thus they should be accompanied by a multiplicative factor of $\Theta(z+ct)$ when represented in the field mode. With these considerations and with the help of the relation $\xi\pm c\,\tau = \pm(c/a)\,\ln{\left[a(z\pm c\,t)/c^2\right]}$, we obtain the field mode expansion
\begin{eqnarray}\label{eq:Rind-FldMode-3p1-2}
        \phi^{R}_{\nu,k_{\perp}} 
&\simeq& \sqrt{\frac{c\,\sinh{(\pi\nu \,c/a)}}{4\pi^4a}} e^{i\,\,\vec{k}_{\perp}.\vec{x}_{\perp}}\,\bigg[   \frac{1}{2}\Gamma\biggl(-\frac{i\,\nu\,c}{a}\biggl)\,\bigg(\frac{k_{\perp}}{2a/c^2}\bigg)^{\frac{i\,\nu\,c}{a}}\theta(z-ct) \,\exp\biggl\{{\frac{i\,\nu c}{a}\ln{\bigg(\frac{a(z-ct)}{c^{2}}\bigg)}}\biggl\}~\nonumber\\
 ~&&~~~~~~~~~~~~~~~~~ +\frac{1}{2}\Gamma\biggl(\frac{i\,\nu\,c}{a}\biggl)\,\bigg(\frac{k_{\perp}}{2a/c^2}\bigg)^{-\frac{i\,\nu\,c}{a}}\theta(z+ct))\, \exp\bigg\{-{\frac{i\,\nu c}{a}\ln{\bigg(\frac{a(z+ct)}{c^{2}}\bigg)}}\bigg\}\bigg]~.
\end{eqnarray}
Then with the help of Eq. \eqref{eq:TP-exc-gen} we obtain the excitation probability as 
\begin{eqnarray}\label{eq:3p1-Ex-atomStatc}
        \mathcal{P}^{ex}_{\nu}(\omega) &=& g^2\,\bigg|\int dt\,\phi_{\nu,k_{\perp}}^{R^{*}}\big(t,z_{0}\big)\,e^{i\,\omega\,t}\bigg|^2~\nonumber\\
        &\simeq&\frac{g^{2}c\,\sinh{(\pi\nu \,c/a)}}{16\,\pi^4a}\bigg| \int_{-\infty}^{\frac{z_{0}}{c}}dt\,e^{i\omega\,t} \Gamma\biggl(\frac{i\nu\,c}{a}\biggl)\,\bigg(\frac{k_{\perp}}{2a/c^2}\bigg)^{-\frac{i\nu\,c}{a}}\,\bigg(\frac{a(z_{0}-ct)}{c^{2}}\biggl)^{-\frac{i\nu c}{a}}\nonumber\\
        ~&&  +\int_{-\frac{z_{0}}{c}}^{\infty}dt\, e^{i\omega\,t}\Gamma\biggl(-\frac{i\nu\,c}{a}\biggl)\,\bigg(\frac{k_{\perp}}{2a/c^2}\bigg)^{\frac{i\nu\,c}{a}}\,\bigg(\frac{a(z_{0}+ct)}{c^{2}}\bigg)^{\frac{i\nu c}{a}}\bigg|^{2} \nonumber\\
        ~&=& \frac{g^{2}c\,\sinh{(\pi\nu \,c/a)}}{16\pi^4a}\,\frac{c^{2}\nu^{2}}{a^{2}\omega^{2}}\biggl|\Gamma\biggl(\frac{i\nu\,c}{a}\biggl)\biggl|^{4}\,e^{-\frac{\pi\nu\,c}{a}} \,\biggl|  \bigg(\frac{2\omega}{k_{\perp}c}\bigg)^{\frac{i\nu\,c}{a}}\,e^{\frac{i\omega\,z_{0}}{c}} + \bigg(\frac{2\omega}{k_{\perp}c}\bigg)^{-\frac{i\nu\,c}{a}}\,e^{\frac{-i\omega\,z_{0}}{c}} \biggl|^{2} \,~\nonumber\\
        ~&=& \frac{g^{2}c}{2\pi^{2}\omega^{2}a}\,\frac{\cos{^{2}\bigg(\frac{\omega\,z_{0}}{c}+\Bar{\varphi}_{ex}\bigg)} }{e^{2\pi\nu\,c/a}-1}~,
\end{eqnarray}
where $\Bar{\varphi}_{ex}=\mathrm{Arg}\big[\{c\,k_{\perp}/(2\omega)\}^{-i\,\nu\,c/a}\big]$. 
We would like to mention that when the atom is sitting at the origin of the Minkowski spacetime, i.e., when $z_{0}=0$, we obtain the excitation probability from Eq. \eqref{eq:3p1-Ex-atomStatc} as
\begin{eqnarray}\label{eq:3p1-Ex-atomStatc-2}
        \mathcal{P}^{ex}_{\nu}(\omega) \simeq \frac{g^{2}c}{2\pi^{2}\omega^{2}a}\,\frac{\cos^{2}{\Bar{\varphi}_{ex}}}{e^{2\pi\nu\,c/a}-1}~.
\end{eqnarray}
Here one can notice that the above expression of the excitation probability will not look similar to the excitation of Eq. \eqref{eq:3p1-atomic-atomAcc-4a-2} corresponding to an accelerated atom when $\omega=\nu$, though the thermal factors become identical. The difference for $\omega=\nu$ arises as the expressions of $\varphi_{ex}$ and $\Bar{\varphi}_{ex}$ are different. Notably $\varphi_{ex}$ depends on the direction of acceleration of the atom, while $\Bar{\varphi}_{ex}$ does not.

\section{Atomic de-excitation without a boundary}\label{sec:AccAtom-DE-WOutBound}
In this section, we study the de-excitation probabilities of atoms without the presence of any boundary in both $(1+1)$ and $(3+1)$ dimensions. In each of these spacetime dimensions, we shall first consider the atom in uniform acceleration seeing the Minkowski vacuum, and then a static atom seeing the Rindler vacuum.

\subsection{De-excitation probability in $(1+1)$ dimensions}
First, we consider the situation where the atom is in uniform acceleration. The de-excitation probability can be obtained like the evaluation of the excitation probability. The expectation is evaluated in the Minkowski vacuum $|0_{M}\rangle$. We consider the expression of Eq. \eqref{eq:TP-deex-gen} to obtain the de-excitation probability
\begin{eqnarray}\label{eq:Pdex-Atom-Acc}
    \mathcal{P}^{de}_{\nu}(\omega)=\frac{g^{2}c}{2\,a\,\nu\,\omega}\frac{1}{1-e^{-2\pi\omega c/a}}~.
\end{eqnarray}
Here we used the mode, mentioned above Eq. (\ref{eq:Pex-Atom-Acc0}).
\vspace{0.15cm}

Second, we consider the situation when the atom is static in the Minkowski background but it responds to the Rindler vacuum $|0_{R}\rangle$. We utilize the same expression of \eqref{eq:TP-deex-gen} to estimate the atomic de-excitation probability (using the mode, introduced above Eq. (\ref{eq:Pex-Atom-Stat0}))
\begin{eqnarray}\label{eq:Pdex-Atom-Stat}
    \mathcal{P}^{de}_{\nu}(\omega)=\frac{g^{2} c}{2\,a\,\omega^2} \frac{1}{1-e^{-2\pi\nu c/a}}~.
\end{eqnarray}
We should note that the expression for de-excitation probability \eqref{eq:Pdex-Atom-Acc} can be obtained from the result corresponding to the excitation probability of Eq. \eqref{eq:Pex-Atom-Acc} by making $\omega\to-\omega$. However, the same is not true for Eqs. \eqref{eq:Pdex-Atom-Stat} and  \eqref{eq:Pex-Atom-Stat}. 
Not only that, it can be checked that by making $\nu\to -\nu$ in Eq. \eqref{eq:Pex-Atom-Stat}, forget about \eqref{eq:Pdex-Atom-Stat}, one cannot get a consistent expression. All these expressions are also provided in Table \ref{tab:Obs1} for the ease of comparison.
Furthermore, we observe that when $\omega=\nu$ the expressions of de-excitations from Eqs. \eqref{eq:Pdex-Atom-Acc} and \eqref{eq:Pdex-Atom-Stat} become the same. Therefore, in $(1+1)$ dimensions and for equal atomic and field frequencies the de-excitation in a uniformly accelerated atom is the same as the de-excitation in a static atom looking into the Rindler vacuum.

\begin{table}[h!]
    \centering
    \begin{tabular}{ ||p{2.0cm}|p{4.0cm}|p{3.5cm}|p{5.5cm}||  }
 \hline
 \multicolumn{4}{|c|}{A tabular list of our observations from \ref{sec:AccAtom-E-WOutBound} and \ref{sec:AccAtom-DE-WOutBound}} \\
 \hline
 Dimensions & Excitation/De-excitation & Atom accelerated/static & Transition probabilities $\mathcal{P}^{ex/de}_{\nu}(\omega)$ \\
 \hline
 \multirow{2}{*}{$\mathbf{(1+1)}$} & Excitation & Accelerated & $\mathlarger{\mathlarger{\mathcal{P}^{ex}_{\nu}(\omega) 
    = \frac{g^{2}c}{2\,a\,\nu\,\omega}\frac{1}{e^{2\pi\omega c/a}-1}}}$ \\\cline{3-4}
 &  & Static & $\mathlarger{\mathlarger{\mathcal{P}^{ex}_{\nu}(\omega)=\frac{g^{2} c}{2\,a\,\omega^2} \frac{1}{e^{2\pi\nu c/a}-1}}}$ \\\cline{2-4}
 \multirow{2}{*}{} & De-excitation & Accelerated & $\mathlarger{\mathlarger{\mathcal{P}^{de}_{\nu}(\omega)=\frac{g^{2}c}{2\,a\,\nu\,\omega}\frac{1}{1-e^{-2\pi\omega c/a}}}}$ \\\cline{3-4}
 &  & Static & $\mathlarger{\mathlarger{\mathcal{P}^{de}_{\nu}(\omega)=\frac{g^{2} c}{2\,a\,\omega^2} \frac{1}{1-e^{-2\pi\nu c/a}}}}$ \\
 \hline
 \multirow{2}{*}{$\mathbf{(3+1)}$} & Excitation & Accelerated & $\mathlarger{\mathlarger{\mathcal{P}^{ex}_{\nu}(\omega) \simeq  \frac{g^{2}c}{2\pi^2\,\nu\, \omega\, a}\, \frac{\cos^{2}{\varphi_{ex}}}{e^{2\pi\omega c/a}-1}}}$ \\\cline{3-4}
 &  & Static & $\mathlarger{\mathlarger{\mathcal{P}^{ex}_{\nu}(\omega) \simeq \frac{g^{2}c}{2\pi^{2}\omega^{2}a}\,\frac{\cos^{2}{\Bar{\varphi}_{ex}}}{e^{2\pi\nu\,c/a}-1}}}$ \\\cline{2-4}
 \multirow{2}{*}{} & De-excitation & Accelerated &  $\mathlarger{\mathlarger{\mathcal{P}^{de}_{\nu}(\omega) \simeq \frac{g^{2}c}{2\pi^2\,\nu\, \omega\, a}\, \frac{\cos^{2}{\varphi_{de}}}{1-e^{-2\pi\omega c/a}}}}$\\\cline{3-4}
 &  & Static & $\mathlarger{\mathlarger{\mathcal{P}^{de}_{\nu}(\omega) \simeq \frac{g^{2}c}{2\pi^{2}\omega^{2}a}\,\frac{\cos^{2}{\Bar{\varphi}_{de}} }{1-e^{-2\pi\nu\,c/a}}}}$ \\
 \hline
\end{tabular}
    \caption{The above table lists our key observations regarding the atomic transitions without the presence of a boundary in $(1+1)$ and $(3+1)$ dimensions. We have listed both the atomic excitation and de-excitation probabilities from Secs. \ref{sec:AccAtom-E-WOutBound} and \ref{sec:AccAtom-DE-WOutBound}.}
    \label{tab:Obs1}
\end{table}


\subsection{De-excitation probability in $(3+1)$ dimensions}
First, we consider that the atom is in uniform acceleration and it responds to the Minkowski field vacuum $|0_{M}\rangle$. The coordinate transformation for the accelerated atom is given in Eq. \eqref{eq:CT-R-2} and the normalized field modes in the Minkowski vacuum are provided in Eq. \eqref{eq:Phi-3p1-WOB}. We utilize these expressions in Eq. \eqref{eq:TP-deex-gen} to obtain the de-excitation probability as
\begin{eqnarray}\label{eq:3p1-DeEx-atomAcc-1}
        \mathcal{P}^{de}_{\nu}(\omega) &=& \frac{g^2}{(2\pi)^3\,2\,\nu}\,\bigg| \frac{2c}{a}\,e^{\frac{\pi\omega\,c}{2a}}\biggl(\frac{1+\cos{\theta}}{1-\cos{\theta}}\biggl)^{-\frac{i\omega\,c}{2a}}\,\mathcal{K}_{-\frac{i\omega\,c}{a}}\Big(\frac{\nu c}{a}\sin{\theta}\Big)\bigg|^2~.
\end{eqnarray}
This expression can be simplified with the asymptotic form of the Bessel function from Eq. \eqref{eq:BesselK-series-exp} for large acceleration, i.e., when $\nu\,c/a\ll 1$. After simplification, we shall get
\begin{eqnarray}\label{eq:3p1-DeEx-atomAcc-2}
        \mathcal{P}^{de}_{\nu}(\omega) &\simeq& \frac{g^{2}c}{2\pi^2\,\nu\, \omega\, a}\, \frac{\cos^{2}{\varphi_{de}}}{1-e^{-2\pi\omega c/a}} ~,
\end{eqnarray}
where $\varphi_{de}=\mathrm{Arg}\Big[\big\{c\nu \sin{\theta} /(2a) \big\}^{i \omega c/a}\Gamma\big(-i\omega c/a\big)\Big]$ is the same as the one given in Eq. \eqref{eq:3p1-atomic-atomAcc-4a-2}. Therefore, one can notice that in Eq. \eqref{eq:3p1-atomic-atomAcc-4a-2} by making $\omega\to -\omega$ one can get Eq. \eqref{eq:3p1-DeEx-atomAcc-2}.

Second, we consider the atom to be static and it responds to the Rindler vacuum $|0_{R}\rangle$. The expressions for normalized field mode solutions are given in Eq. \eqref{eq:Rind-FldMode-3p1} and we further consider the field mode expansion given in \eqref{eq:Rind-FldMode-3p1-2} with Minkowski coordinate transformation to $(t,\,z)$. If we consider the atom to be static at $z=0$, with the help of Eq. \eqref{eq:TP-deex-gen} we can obtain the de-excitation probability to be
\begin{eqnarray}\label{eq:3p1-DeEx-atomStatc}
        \mathcal{P}^{de}_{\nu}(\omega) \simeq \frac{g^{2}c}{2\pi^{2}\omega^{2}a}\,\frac{\cos^{2}{\Bar{\varphi}_{de}} }{1-e^{-2\pi\nu\,c/a}}~,
\end{eqnarray}
where $\Bar{\varphi}_{de}= \mathrm{Arg}[\{c\,k_{\perp} /(2\omega)\}^{ i\,\nu\,c/a}]$. We notice that when the atom is accelerated, one can get the de-excitation probability of Eq. \eqref{eq:3p1-DeEx-atomAcc-2} by making $\omega\to-\omega$ in the excitation probability \eqref{eq:3p1-atomic-atomAcc-4a-2}. We also observe that when the atom is static the de-excitation probability \eqref{eq:3p1-DeEx-atomStatc} cannot be obtained from the excitation probability \eqref{eq:3p1-Ex-atomStatc-2}, by changing the sign of the atomic or the field frequencies. Both of these observations are similar to the $(1+1)$ dimensional scenario. Our observations also suggest that the above $(3+1)$ dimensional de-excitation probabilities do not match when $\omega=\nu$, contrary to the $(1+1)$ dimensional case. We infer that, for an accelerated atom in the $(3+1)$ dimension, the directional dependence of the transition probability plays a crucial role in this mismatch. All these expressions of atomic excitation and de-excitation probabilities without the presence of a boundary are presented in Table \ref{tab:Obs1} for the ease of comparison.

\section{Atomic excitation in the presence of a mirror}\label{sec:AccAtom-E-WBound}
In this section, we will investigate the excitation probability of a uniformly accelerated or static atom in the presence of a static or uniformly accelerated mirror. We will recall the results corresponding to a $(1+1)$ dimensional scenario from \cite{Svidzinsky:2018jkp}. Then we will provide the expressions for the excitation probabilities for similar scenarios in $(3+1)$ dimensions. As we will see the computations in $(3+1)$ dimensions may result in observations that depend on the specific direction of acceleration.

\subsection{Excitation probability in $(1+1)$ dimensions}

The virtual transition in a static atom in the presence of a uniformly accelerated mirror is presented in \cite{Svidzinsky:2018jkp}. This work also discusses the excitation of a uniformly accelerated atom in the presence of a static reflecting mirror. It should be mentioned that both of these works dealt with a set-up in $(1+1)$ dimensional flat spacetime. Here we shall not re-derive the expressions for the transition probabilities, rather quote the expression obtained in \cite{Svidzinsky:2018jkp}. For instance, when the atom is accelerated and the mirror is static the excitation probability is given by
\begin{eqnarray}\label{eq:ExProb-1p1-AccAtm}
    \mathcal{P}^{ex}_{\nu}(\omega) = \frac{2\,c\,g^2}{a\,\omega\,\nu}\,\frac{\sin^2{(\nu\,z_{0}/c+\varphi_{1})}}{e^{2\pi\,\omega\,c/a}-1}~.
\end{eqnarray}
Where the mirror is stationed at a distance $z_{0}$ along the $z$ axis, and we have $\varphi_{1}=\mathrm{Arg}[(c\,\nu/a)^{i\,c\,\omega/a}\,\Gamma(-i\,c\,\omega/a)]$.\\

At the same time, if the atom is kept static at $z=z_{0}$ and the mirror is in uniform acceleration, the virtual excitation probability of the atom can be found to be
\begin{eqnarray}\label{eq:ExProb-1p1-AccMirr}
    \mathcal{P}^{ex}_{\nu}(\omega) = \frac{2\,c\,g^2}{a\,\omega^2}\,\frac{\sin^2{(\omega\,z_{0}/c+\varphi_{2})}}{e^{2\pi\,\nu\,c/a}-1}~.
\end{eqnarray}
In the above expression the angle $\varphi_{2}$ is given by $\varphi_{2}=\mathrm{Arg}[(c\,\omega/a)^{i\,c\,\nu/a}\,\Gamma(-i\,c\,\nu/a)]$. From the above expressions we observe that, in $(1+1)$ dimensions and in the presence of a mirror, the atomic excitation probabilities are the same when $\omega=\nu$. Thus in terms of equivalence, this scenario is similar to the one without a boundary. We would also like to mention that the above expressions from Eqs. \eqref{eq:ExProb-1p1-AccAtm} and \eqref{eq:ExProb-1p1-AccMirr} are slightly different in the pre-factors compared to the expressions from \cite{Svidzinsky:2018jkp}, as we have utilized the normalized field mode solutions to obtain these results.

\subsection{Excitation probability in $(3+1)$ dimensions}
In this part of the section, we consider the $(3+1)$ dimensional scenario and investigate the atomic excitation probability in the presence of an infinite reflecting mirror. First, we will consider the situation when the atom is in uniform acceleration and the mirror is static. Second, we shall consider the atom to be static and the mirror is in uniform acceleration.

\subsubsection{When the atom is in uniform acceleration}

Here we consider the atom in uniform acceleration along the $z$ direction and the mirror is kept static at $z=z_{0}$, on the $x-y$ plane. Eq. \eqref{eq:CT-R-2} gives the coordinate transformation relevant to the atom. In $(3+1)$ dimensional Minkowski spacetime the normalized field mode solution for a neutral photon field is given by 
\begin{eqnarray}\label{eq:Phi-3p1-Atom-Mirror}
    \phi_{\nu}[t(\tau),z(\tau)] = \frac{1}{\sqrt{(2\pi)^3\,2\,\nu}}e^{-i\,\nu\,t+i\,\Vec{k}_{\perp}.\Vec{x}_{\perp}}\Big[e^{-i\,k_{z}(z-z_{0})}-e^{i\,k_{z}(z-z_{0})}\Big]~.
\end{eqnarray}
We would like to mention that this expression for the field mode solution is inspired by the expression of $(1+1)$ dimensional field mode in the presence of a mirror as obtained in \cite{Svidzinsky:2018jkp, book:Birrell}. Here we obtained this field mode by considering two positive frequency modes, as mentioned in Eq. \eqref{eq:Phi-3p1-WOB}, with one mode propagating in the forward $z$-direction and another in the backward $z$-direction. One can notice that with this arrangement, used in \cite{Svidzinsky:2018jkp} as well, the field mode vanishes when $z=z_{0}$, i.e., at the position of the mirror. We should also mention that this mode propagates in the domain $z\in (-\infty,z_{0}]$ and is properly normalized.
We utilize this field mode solution along with the expression from Eq. \eqref{eq:TP-exc-gen} to obtain the excitation probability. In particular, in this scenario, the expression of this excitation probability, which also corresponds to the simultaneous emission of a photon, is
\begin{eqnarray}\label{eq:ExProb-3p1-AccAtm-1}
    \mathcal{P}^{ex}_{\nu}(\omega) &=& \frac{g^2}{(2\pi)^3\,2\nu}\,\bigg|\int d\tau\,e^{i\,\nu\,t+i\,\omega\,\tau}\Big[e^{i\,k_{z}(z-z_{0})}-e^{-i\,k_{z}(z-z_{0})}\Big]\bigg|^2~\nonumber\\
    ~&=& \frac{g^2}{(2\pi)^3\,2\nu}\,\bigg|\int_{-\infty}^{\infty} d\tau\,\Big[\exp{\Big\{\frac{i\,k_{z}\,c^2}{a}\cosh{\Big(\frac{a\,\tau}{c}\Big)}-i\,k_{z}\,z_{0}\Big\}}~\exp{\Big\{\frac{i\,\nu\,c}{a}\sinh{\Big(\frac{a\,\tau}{c}\Big)}+i\,\omega\,\tau\Big\}}-c.c.\Big]\bigg|^2~.
\end{eqnarray}
We consider a change of variables $x=e^{a\,\tau/c}$. Then the excitation probability becomes 
\begin{eqnarray}\label{eq:ExProb-3p1-AccAtm-2}
    \mathcal{P}^{ex}_{\nu}(\omega) &=& \frac{c^2\,g^2}{(2\pi)^3\,2\,\nu\,a^2}\,\bigg|\int_{-\infty}^{\infty} dx\,\Big[e^{-i\,k_{z}\,z_{0}}\,\exp{\Big\{\frac{i\,k_{z}\,c^2}{2\,a}\Big(x+\frac{1}{x}\Big)\Big\}}~\exp{\Big\{\frac{i\,\nu\,c}{2\,a}\Big(x-\frac{1}{x}\Big)\Big\}}\,x^{-1+i\,\omega\,c/a}-c.c.\Big]\bigg|^2~\nonumber\\
    ~&=& \frac{c^2\,g^2}{(2\pi)^3\,2\,\nu\,a^2}\,\bigg|\pi\,e^{-i\,k_{z}\,z_{0}}\,\Big(\frac{\nu-c\,k_{z}}{\nu+c\,k_{z}}\Big)^{\frac{i\,\omega\,c}{2\,a}}\,\mathbb{Y}\Big[-\frac{i\,\omega\,c}{a},-\frac{c\sqrt{\nu^2-k_{z}^2\,c^2}}{a}\Big]-c.c.\bigg|^2~.
\end{eqnarray}
In the above expression, $\mathbb{Y}[\mu,\,z]\equiv \mathbb{Y}_{\mu}(z)$ denotes the Bessel function of the second kind of order $\mu$. For a general acceleration $a$ of the atom, this expression cannot be expressed in a more simplified form. However, for large acceleration $a$, this expression can be simplified to express a Planckian distribution with a sinusoidal periodicity. For instance, for a small argument $(z\ll 1)$, one can asymptotically expand the Bessel function $\mathbb{Y}_{\mu}(z)\sim -[(z/2)^{\mu}\,\cos{(\pi\,\mu)}\,\Gamma(-\mu)+(z/2)^{-\mu}\,\Gamma(\mu)]/\pi$, see the limiting form of Bessel function $(10.7.6)$ of \cite{NIST:DLMF}. If we utilize this expression with the assumption $\omega\,c/a\ll 1$, we shall get the above transition probability as $\mathcal{P}=|\mathcal{A}_{+}-\mathcal{A}_{-}|\,g^2/\big\{(2\pi)^3\,2\nu\big\}$, with $\mathcal{A}_{\pm}$ given by
\begin{eqnarray}\label{eq:ExProb-3p1-AccAtm-3}
    \mathcal{A}_{\pm} &=& \frac{c}{a}e^{\mp \,i\,k_{z}\,z_{0}} \bigg[ \left\{-\frac{i\, c}{2a}(\nu\pm c\,k_{z})\right\}^{-i\,\omega\,c/a}\,\Gamma\left(\frac{i\,\omega\,c}{a}\right) + \left\{\frac{i\, c}{2a}(\nu\mp c\,k_{z})\right\}^{i\,\omega\,c/a}\,\Gamma\left(-\frac{i\,\omega\,c}{a}\right)\bigg]~.
\end{eqnarray}
Therefore for large acceleration $a$ and mathematically when $\omega\,c/a\ll 1$, one can express the excitation probability as 
\begin{eqnarray}\label{eq:ExProb-3p1-AccAtm-4}
    \mathcal{P}^{ex}_{\nu}(\omega) &\simeq& \frac{c\,g^2}{2\,\pi^2\nu\,\omega\,a}~\frac{1}{e^{2\pi\omega\,c/a}-1}~|\sin{(k_{z}\,z_{0}-\psi_{1})}+\sin{(k_{z}\,z_{0}-\psi_{2})}|^2~.
\end{eqnarray}
Here we denote $\psi_{j}=\mathrm{Arg}[\kappa_{j}]$ where $j$ can be either $1$ or $2$ and $\mathrm{Arg}[\kappa_{j}]=tan^{-1}\{\mathrm{Im}[\kappa_{j}]/\mathrm{Re}[\kappa_{j}]\}$. The expressions of $\kappa_{1}$ and $\kappa_{2}$ are given by $\kappa_{1}=\Gamma(i\,\omega\,c/a)\,\{c(\nu+c\,k_{z})/(2a)\}^{-i\,\omega\,c/a}$ and $\kappa_{2}=\Gamma(-i\,\omega\,c/a)\,\{c(\nu-c\,k_{z})/(2a)\}^{i\,\omega\,c/a}$. It is to be noted that for large $a/(c\,\omega)$ the quantities $\kappa_{1}$ and $\kappa_{2}$ behave as $\kappa_{1}\simeq -i\,a/(c\,\omega)-\log \left\{c\, (c\, k_{z}+\nu )/(2\,a)\right\}-\gamma$ and $\kappa_{2}\simeq i\,a/(c\,\omega)-\log \left\{c\, (\nu-c\, k_{z})/(2\,a)\right\}-\gamma$, where $\gamma$ denotes the Euler–Mascheroni constant. Thus for large acceleration, the values of $\psi_{j}$ are respectively $\psi_{1}=-\pi/2-(c\,\omega/a)\big[\log \left\{c\, (c\, k_{z}+\nu )/(2\,a)\right\}+\gamma\big]$ and $\psi_{2}=\pi/2+(c\,\omega/a)\big[\log \left\{c\, (\nu-c\, k_{z})/(2\,a)\right\}+\gamma\big]$. In the limit of large acceleration, if we put these values we shall get the excitation probability
\begin{eqnarray}\label{eq:ExProb-3p1-AccAtm-5}
 \mathcal{P}^{ex}_{\nu}(\omega) &\simeq& \frac{2\,c\, g^{2}}{\pi^{2}\omega\,\nu\, a}~\frac{\cos^{2}{(k_{z}z_{0})}}{e^{2\pi\omega\,c/a}-1}\,\sin^{2}{\left(\psi_{0}\right)}~.
\end{eqnarray}
In the last expression $\psi_{0}\sim\mathcal{O}(1/a^2)$. From the above expressions, we observe that the excitation probability for an accelerated atom in the presence of a mirror is dependent on the direction along which the atom is accelerated. In particular, our observations suggest that this directional dependence is present in both the general acceleration \eqref{eq:ExProb-3p1-AccAtm-2} and large acceleration \eqref{eq:ExProb-3p1-AccAtm-5} scenarios.

\subsubsection{When the mirror is in uniform acceleration}

Here we consider the atom to be static at $z=z_{0}$ and the mirror is in uniform acceleration. In this scenario, the field mode solutions should be estimated in the mirror's frame, i.e., in the Rindler frames. In this regard, we consider the Minkowski line element in Rindler coordinates as described in Eq. \eqref{eq:Rindler-LineEl}. The normalized field mode solutions can be obtained from Eq. \eqref{eq:Rind-FldMode-3p1}. We consider the large acceleration scenario, and $|a\,\xi/c^2|<1$. This last inequality corresponds to the mirror trajectory being bound by the Rindler horizons, which is generally the case for any observer in uniform acceleration. For a field mode moving along the $+\xi$ direction and for large acceleration, one can series expand the modified Bessel function in it as (see Eq. \eqref{eq:BesselK-series-exp})
\begin{eqnarray}\label{eq:ExProb-3p1-AccMirr-BsslK-asymp1}
    \mathcal{K}_{i\nu c/a}\bigg(\frac{k_{\perp}e^{a\,\xi/c^2}}{a/c^2}\bigg) &\simeq& \frac{1}{2}\Gamma\Big(\frac{i\,\nu\,c}{a}\Big)\,\bigg(\frac{k_{\perp}e^{a\,\xi/c^2}}{2a/c^2}\bigg)^{-\frac{i\,\nu\,c}{a}}~\nonumber\\
 ~&& ~~+~   \frac{1}{2}\Gamma\Big(-\frac{i\,\nu\,c}{a}\Big)\,\bigg(\frac{k_{\perp}e^{a\,\xi/c^2}}{2a/c^2}\bigg)^{\frac{i\,\nu\,c}{a}}~.
\end{eqnarray}
At the same time, for modes along $-\xi$ direction, we have for large acceleration
\begin{eqnarray}\label{eq:ExProb-3p1-AccMirr-BsslK-asymp2}
    \mathcal{K}_{i\nu c/a}\bigg(\frac{k_{\perp}e^{-a\,\xi/c^2}}{a/c^2}\bigg) &\simeq& \frac{1}{2}\Gamma\Big(\frac{i\,\nu\,c}{a}\Big)\,\bigg(\frac{k_{\perp}e^{-a\,\xi/c^2}}{2a/c^2}\bigg)^{-\frac{i\,\nu\,c}{a}}~\nonumber\\
 ~&& ~~+~   \frac{1}{2}\Gamma\Big(-\frac{i\,\nu\,c}{a}\Big)\,\bigg(\frac{k_{\perp}e^{-a\,\xi/c^2}}{2a/c^2}\bigg)^{\frac{i\,\nu\,c}{a}}~.
\end{eqnarray}
On the other hand, the Minkowski coordinates and the coordinates in the Mirror's frame have a relation $\xi\pm c\,\tau = \pm(c/a)\,\ln{\left[a(z\pm c\,t)/c^2\right]}$, see Ref. \cite{Svidzinsky:2018jkp}. This expression can also be obtained with the help of the Minkowski to Rindler coordinate transformation from Eq. \eqref{eq:CT-R}. Similar to \cite{Svidzinsky:2018jkp}, here also the mode function $\phi_{\nu,\,k_{\perp}}(t,z)$ in the mirror's frame can be obtained by plugging the expressions of $\xi\pm c\,\tau$ in $\phi_{\nu,\,k_{\perp}}(t,z)=\big(\phi^{R}_{\nu,k_{\perp}}(\tau,\xi,\Vec{x}_{\perp})-\phi^{R}_{\nu,k_{\perp}}(\tau,-\xi,\Vec{x}_{\perp})\big)$ with the proper consideration of the right and left moving field modes, where $\phi^{R}_{\nu,k_{\perp}}(\tau,\xi,\Vec{x}_{\perp})$ is given by Eq. \eqref{eq:Rind-FldMode-3p1} corresponding to a $(3+1)$ dimensional scenario. One can notice that a crucial part of $\phi^{R}_{\nu,k_{\perp}}(\tau,\xi,\Vec{x}_{\perp})$, that determines the left and right moving nature of the mode, is series expanded in Eq. \eqref{eq:ExProb-3p1-AccMirr-BsslK-asymp1} for large acceleration. From Eq. \eqref{eq:ExProb-3p1-AccMirr-BsslK-asymp1} we further observe that this expansion has both right and left moving parts, e.g., $e^{i\,\nu\,\xi/c}$ is right moving and $e^{-i\,\nu\,\xi/c}$ is left moving. Similarly in $\phi^{R}_{\nu,k_{\perp}}(\tau,-\xi,\Vec{x}_{\perp})$ also there are two parts, one right and another left moving. 
Next, we consider the procedure as elaborated in the discussion below Eq. \eqref{eq:Rind-FldMode-3p1}, i.e., the right and left moving parts are accompanied by $\Theta(z-ct)$ and $\Theta(z+ct)$ when constructing the full photon mode. Then the mode functions in the Mirror's frame, with the help of Eqs. \eqref{eq:ExProb-3p1-AccMirr-BsslK-asymp1} and \eqref{eq:ExProb-3p1-AccMirr-BsslK-asymp2}, in terms of the Minkowski coordinates will become
\begin{eqnarray}\label{eq:ExProb-3p1-AccMirr-Phi}
    \phi_{\nu,\,k_{\perp}}(t,z) &=& \sqrt{\frac{c\,\sinh{(\pi\,\nu\,c/a)}}{4\pi^4a}}\,
\frac{e^{i\,\Vec{k}_{\perp}.\Vec{x}_{\perp}}}{2}\,\nonumber\\
~&\times&\bigg[\Theta(z-c\,t)\, \Gamma\Big(-\frac{i\,\nu\,c}{a}\Big)\,\bigg(\frac{k_{\perp}}{2a/c^2}\bigg)^{\frac{i\,\nu\,c}{a}}\, \exp{\Big[\frac{i\,\nu\,c}{a}\,\ln{\Big\{\frac{a}{c^2}(z-c\,t)\Big\}}\Big]} \nonumber\\
~&+& \Theta(z+c\,t)\,\Gamma\Big(\frac{i\,\nu\,c}{a}\Big)\,\bigg(\frac{k_{\perp}}{2a/c^2}\bigg)^{-\frac{i\,\nu\,c}{a}}\, \exp{\Big[-\frac{i\,\nu\,c}{a}\,\ln{\Big\{\frac{a}{c^2}(z+c\,t)\Big\}}\Big]}\nonumber\\
~&-&\Theta(z+c\,t)\, \Gamma\Big(-\frac{i\,\nu\,c}{a}\Big)\,\bigg(\frac{k_{\perp}}{2a/c^2}\bigg)^{\frac{i\,\nu\,c}{a}}\, \exp{\Big[-\frac{i\,\nu\,c}{a}\,\ln{\Big\{\frac{a}{c^2}(z+c\,t)\Big\}}\Big]} \nonumber\\
~&-& \Theta(z-c\,t)\,\Gamma\Big(\frac{i\,\nu\,c}{a}\Big)\,\bigg(\frac{k_{\perp}}{2a/c^2}\bigg)^{-\frac{i\,\nu\,c}{a}}\, \exp{\Big[\frac{i\,\nu\,c}{a}\,\ln{\Big\{\frac{a}{c^2}(z-c\,t)\Big\}}\Big]}\bigg]~.
\end{eqnarray}
Similar construction for $(1+1)$ dimensions has been done in \cite{Svidzinsky:2018jkp}.
The excitation probability of a static atom getting excited due to the presence of the accelerating mirror is 
\begin{subequations}\label{eq:ExProb-3p1-AccMirr-1}
\begin{eqnarray}
    \mathcal{P}^{ex}_{\nu}(\omega) &=& \frac{g^2\,c\,\sinh{(\pi\,\nu\,c/a)}}{16\pi^4\,a}~\nonumber\\
    ~&\times&\bigg| \Gamma\Big(\frac{i\,\nu\,c}{a}\Big)\,\bigg(\frac{k_{\perp}}{2a/c^2}\bigg)^{-\frac{i\,\nu\,c}{a}}\, \int_{-\infty}^{z_{0}/c}dt\,\exp{\Big[-\frac{i\,\nu\,c}{a}\,\ln{\Big\{\frac{a}{c^2}(z-c\,t)\Big\}}+i\omega\,t\Big]} \nonumber\\
~&+& \Gamma\Big(-\frac{i\,\nu\,c}{a}\Big)\,\bigg(\frac{k_{\perp}}{2a/c^2}\bigg)^{\frac{i\,\nu\,c}{a}}\, \int_{-z_{0}/c}^{\infty}dt\, \exp{\Big[\frac{i\,\nu\,c}{a}\,\ln{\Big\{\frac{a}{c^2}(z+c\,t)\Big\}}+i\omega\,t\Big]}\nonumber\\
~&-& \Gamma\Big(\frac{i\,\nu\,c}{a}\Big)\,\bigg(\frac{k_{\perp}}{2a/c^2}\bigg)^{-\frac{i\,\nu\,c}{a}}\, \int_{-z_{0}/c}^{\infty}dt\, \exp{\Big[\frac{i\,\nu\,c}{a}\,\ln{\Big\{\frac{a}{c^2}(z+c\,t)\Big\}}+i\omega\,t\Big]} \nonumber\\
~&-& \Gamma\Big(-\frac{i\,\nu\,c}{a}\Big)\,\bigg(\frac{k_{\perp}}{2a/c^2}\bigg)^{\frac{i\,\nu\,c}{a}}\, \int_{-\infty}^{z_{0}/c}dt\,\exp{\Big[-\frac{i\,\nu\,c}{a}\,\ln{\Big\{\frac{a}{c^2}(z-c\,t)\Big\}}+i\omega\,t\Big]}\bigg|^2~.\label{eq:3p1-atomic-mirrorAcc-8b}
\end{eqnarray}
\end{subequations}
In the above expression, we consider a change of variables $x_{\pm}=\omega(t\pm z_{0}/c)$. Then the above expression is simplified to
\begin{eqnarray}\label{eq:ExProb-3p1-AccMirr-2}
    \mathcal{P}^{ex}_{\nu}(\omega) &=& \frac{g^2\,c\,\sinh{(\pi\,\nu\,c/a)}}{4\pi^4\,a\,\omega^2}~\mathrm{Im}\bigg[\Gamma\Big(\frac{i\,\nu\,c}{a}\Big)\,\bigg(\frac{k_{\perp}c^2}{2a}\bigg)^{-\frac{i\,\nu\,c}{a}}\bigg]^2\,\nonumber\\
    ~&\times&\bigg| e^{\frac{i\omega z_{0}}{c}} \int_{-\infty}^{0}dx_{-}~\Big(-\frac{a\,x_{-}}{\omega\,c}\Big)^{-\frac{i\nu\,c}{a}}\,e^{i\,x_{-}}-e^{-\frac{i\omega z_{0}}{c}} \int_{0}^{\infty}dx_{+}~\Big(\frac{a\,x_{+}}{\omega\,c}\Big)^{\frac{i\nu\,c}{a}}\,e^{i\,x_{+}}\bigg|^2 ~.
\end{eqnarray}
Next we consider a change of variables $x_{-}\to (-x_{-})$, and the above integral becomes
\begin{eqnarray}\label{eq:ExProb-3p1-AccMirr-3}
    \mathcal{P}^{ex}_{\nu}(\omega) &=& \frac{g^2\,c\,\sinh{(\pi\,\nu\,c/a)}}{4\pi^4\,a\,\omega^2}~\mathrm{Im}\bigg[\Gamma\Big(\frac{i\,\nu\,c}{a}\Big)\,\bigg(\frac{k_{\perp}c^2}{2a}\bigg)^{-\frac{i\,\nu\,c}{a}}\bigg]^2\,\nonumber\\
    ~&\times&\bigg| e^{\frac{i\omega z_{0}}{c}} \int_{0}^{\infty}dx_{-}~\Big(\frac{a\,x_{-}}{\omega\,c}\Big)^{-\frac{i\nu\,c}{a}}\,e^{-i\,x_{-}}-e^{-\frac{i\omega z_{0}}{c}} \int_{0}^{\infty}dx_{+}~\Big(\frac{a\,x_{+}}{\omega\,c}\Big)^{\frac{i\nu\,c}{a}}\,e^{i\,x_{+}}\bigg|^2 \nonumber\\
    ~&=&\frac{g^2\,c\,\sinh{(\pi\,\nu\,c/a)}}{4\pi^4\,a\,\omega^2}~\mathrm{Im}\bigg[\Gamma\Big(\frac{i\,\nu\,c}{a}\Big)\,\bigg(\frac{k_{\perp}c^2}{2a}\bigg)^{-\frac{i\,\nu\,c}{a}}\bigg]^2~
    \mathrm{Im}\bigg[e^{\frac{i\omega z_{0}}{c}}\,\Gamma\Big(-\frac{i\,\nu\,c}{a}\Big)\,\bigg(\frac{a}{\omega\,c}\bigg)^{-\frac{i\,\nu\,c}{a}}\bigg]^2\,\frac{\nu^2\,c^2\,e^{-\pi\,\nu\,c/a}}{a^2}\nonumber\\
    ~&=& \frac{2\,g^2\,c\,\sin^2{(\Bar{\psi}_{1})}}{\pi^2\,a\,\omega^2}~\frac{\sin^2{(\Bar{\psi}_{2}+\omega\,z_{0}/c)}}{e^{2\,\pi\,\nu\,c/a}-1}~.
\end{eqnarray}
In the above expression, the quantities $\Bar{\psi}_{1}$ and $\Bar{\psi}_{2}$ are given by $\Bar{\psi}_{1}=\mathrm{Arg}[\Bar{\kappa}_{1}]$ and $\Bar{\psi}_{2}=\mathrm{Arg}[\Bar{\kappa}_{2}]$, with $\Bar{\kappa}_{1}=\Gamma\big(i\,\nu\,c/a\big)\,\big\{k_{\perp}c^2/(2a)\big\}^{-i\,\nu\,c/a}$ and $\Bar{\kappa}_{2}=\Gamma\big(-i\,\nu\,c/a\big)\,\big\{a/(\omega\,c)\big\}^{-i\,\nu\,c/a}$.
When $a\to\infty$ we get $\Bar{\kappa}_{1}\sim -i\,a/(c\,\nu)$ and $\Bar{\kappa}_{2}\sim i\,a/(c\,\nu)$. Then in this limit, we shall have $\Bar{\psi}_{1}=-\pi/2$ and $\Bar{\psi}_{2}=\pi/2$, and the above atomic excitation probability becomes
\begin{eqnarray}\label{eq:ExProb-3p1-AccMirr-4}
    \mathcal{P}^{ex}_{\nu}(\omega) &\simeq& \frac{2\,g^2\,c}{\pi^2\,a\,\omega^2}\, \frac{\cos^2{\Big(\frac{\omega\,z_{0}}{c}\Big)}}{e^{\frac{2\pi\,\nu\,c}{a}}-1} ~.
\end{eqnarray}
On the other hand, if we keep $\mathcal{O}(1/a^2)$ terms we shall get
\begin{eqnarray}\label{eq:ExProb-3p1-AccMirr-5}
    \mathcal{P}^{ex}_{\nu}(\omega) &\simeq& \frac{2\,g^2\,c}{\pi^2\,a\,\omega^2}~\cos^{2}{\left(\frac{1}{\frac{a}{\gamma\nu\,c}-\frac{1}{2}\left(\gamma^{2}+\frac{\pi^{2}}{6}\right)\frac{\nu\,c}{\gamma\,a}}\right)} \times\frac{\cos^{2}{\left(\frac{\omega\,z_{0}}{c}+\frac{1}{\frac{a}{\gamma\nu\,c}-\frac{1}{2}\left(\gamma^{2}+\frac{\pi^{2}}{6}\right)\frac{\nu\,c}{\gamma\,a}}\right)}}{e^{2\,\pi\,\nu\,c/a}-1}~.
\end{eqnarray}
From Eqs. \eqref{eq:ExProb-3p1-AccMirr-4} and \eqref{eq:ExProb-3p1-AccMirr-5} we observe that in a large acceleration scenario, the excitation probability of a static atom in the presence of an accelerated mirror becomes independent of the direction along which the mirror is accelerated. This is contrary to the situation when the atom is accelerated. We also observe that because of this particular feature, the excitation probabilities from the two situations, corresponding to one with accelerated atom and static mirror and the other with static atom and accelerated mirror, do not completely match when the atomic and the field frequencies become equal, see Eqs. \eqref{eq:ExProb-3p1-AccAtm-5} and \eqref{eq:ExProb-3p1-AccMirr-4}. However, it should also be noted that their thermal factors do match when $\omega=\nu$.

\section{Atomic de-excitation in the presence of a mirror}\label{sec:AccAtom-DE-WBound}
In this section, we study the atomic de-excitation in the presence of a reflecting mirror. Although, it may seem that compared to atomic excitation in the de-excitation the transition probability should have a negative atomic frequency $(\omega)$, we will see that this is not the case at all scenarios. In the subsequent analysis, we shall consider the $(1+1)$ and $(3+1)$ dimensional scenarios separately, and consider two specific atom-mirror set-ups which were also taken in the last section. In one of these set-ups, the atom is in uniform acceleration and the mirror is static, and in the other the atom is static and the mirror moves in uniform acceleration.

\subsection{De-excitation in $(1+1)$ dimensions}
In this part of the section, we consider the $(1+1)$ dimensional scenario and first assume a set-up where the atom is in uniform acceleration and the mirror is static at $z=z_{0}$. 
The coordinate transformation relevant to the atom is given by Eq. \eqref{eq:CT-R-2} and the field mode solution is given \cite{Svidzinsky:2018jkp, book:Birrell} by
\begin{eqnarray}\label{eq:Phi-Dex-1p1-AccAtm}
    \phi_{\nu} =\frac{1}{\sqrt{4\pi\,\nu}} e^{-i\,\nu\,t}\,\Big[e^{-i\,k(z-z_{0})}-e^{i\,k(z-z_{0})}\Big]~.
\end{eqnarray}
For an accelerated atom we also have $t\pm z/c=\pm (c/a)\,e^{\pm a\,\tau/c}$, see Eq. \eqref{eq:CT-R-2}. With the help of the general expression for the de-excitation probability from Eq. \eqref{eq:TP-deex-gen} we obtain the de-excitation probability in a uniformly accelerated atom in the presence of a static mirror as
\begin{eqnarray}\label{eq:DexProb-1p1-AccAtm}
    \mathcal{P}^{de}_{\nu}(\omega) &=& g^2\bigg|\int_{-\infty}^{\infty} d\tau\,\Big[e^{-i\,\omega\,\eta}\,e^{-i\,k\,z_{0}}\,\exp{\big\{-i\,\nu\,c\,e^{-a\,\tau/c}/a\big\}}-c.c.\Big]\bigg|^2~\nonumber\\
    ~&=& \frac{2\,c\,g^2}{a\,\omega\,\nu}\,\frac{\sin^2{(\nu\,z_{0}/c+\Bar{\varphi}_{1})}}{1-e^{-2\pi\,c\,\omega/a}}~.
\end{eqnarray}
In the above expression we have $\Bar{\varphi}_{1}=\mathrm{Arg}[(c\,\nu/a)^{-i\,\omega\,c/a}\,\Gamma(i\,\omega\,c/a)]$. One can notice that this expression is the same as the expression of Eq. \eqref{eq:ExProb-1p1-AccAtm} with $\omega\to -\omega$. Therefore, our observations suggest that in $(1+1)$ dimensions the excitation and de-excitation probabilities in a uniformly accelerated atom in the presence of a reflecting mirror are related among themselves via a sign change in the atomic frequency $\omega$.\\

Next, we consider the situation when the atom is static at $z=z_{0}$ and the mirror is in uniform acceleration. In this scenario, with the help of Eq. \eqref{eq:TP-deex-gen} and the field mode solutions provided in \cite{Svidzinsky:2018jkp} we get the atomic de-excitation probability as
\begin{eqnarray}\label{eq:DexProb-1p1-AccMirr-1}
    \mathcal{P}^{de}_{\nu}(\omega) &=& \frac{g^2}{4\pi\,\nu}\bigg|\int_{-\infty}^{z_{0}/c} dt\,\exp{\Big\{-\frac{i\,\nu\,c}{a}\ln{\big[\frac{a}{c^2}(z_{0}-c\,t)\big]}-i\,\omega\,t\Big\}}-\int_{-z_{0}/c}^{\infty} dt\,\exp{\Big\{\frac{i\,\nu\,c}{a}\ln{\big[\frac{a}{c^2}(z_{0}+c\,t)\big]}-i\,\omega\,t\Big\}}\bigg|^2~.
\end{eqnarray}
We consider change of variables of $x=z_{0}-c\,t$ and $\bar{x}=z_{0}+c\,t$. Then the above integral becomes
\begin{eqnarray}\label{eq:DexProb-1p1-AccMirr-2}
    \mathcal{P}^{de}_{\nu}(\omega) &=& \frac{g^2}{4\pi\,\nu\,c} \bigg|\int_{0}^{\infty}dx\,\Big(\frac{a\,x}{c}\Big)^{-i\,\nu\,c/a}\,e^{-i\omega(z_{0}-x)/c}-\int_{0}^{\infty}d\bar{x}\,\Big(\frac{a\,\bar{x}}{c}\Big)^{i\,\nu\,c/a}\,e^{i\omega(z_{0}-\bar{x})/c}\bigg|^2\nonumber\\
    &=& \frac{2\,c\,g^2}{\omega^2\,a}\,\frac{\sin^2{(\omega\,z_{0}/c+\Bar{\varphi}_{2})}}{1-e^{-2\pi\,c\,\nu/a}}~.
\end{eqnarray}
In the above expression $\Bar{\varphi}_{2}=\mathrm{Arg}[(c\,\omega/a)^{-i\,\nu\,c/a}\,\Gamma(i\,\nu\,c/a)]$. Notice that one can obtain this expression from the excitation probability's expression \eqref{eq:ExProb-1p1-AccMirr} by changing the sign of the field frequency, i.e., by making $\nu\to -\nu$. This is in contrast to the situation when there is no reflecting boundary. Therefore, in $(1+1)$ dimensions when the atom is static and the mirror is in uniform acceleration, we observe that the atomic excitation and de-excitation probabilities are related among themselves via a sign change in the field frequency $\nu$. Moreover, we also observe that when $\omega=\nu$ the de-excitation probabilities corresponding to two different scenarios, one with an accelerated atom and another with a static atom, become equal, see Eqs. \eqref{eq:DexProb-1p1-AccAtm} and \eqref{eq:DexProb-1p1-AccMirr-2}.

\subsection{De-excitation in $(3+1)$ dimensions}

In this part of the section, we consider the $(3+1)$ dimensional scenario of atomic de-excitation in the presence of a mirror. First, we shall look into the case where the atom is in uniform acceleration and the mirror is static. Then we shall investigate the case where the atom is static and the mirror is in uniform acceleration.

\subsubsection{When the atom is in uniform acceleration}

Here we consider the case when the atom is in uniform acceleration and the mirror is static at $z=z_{0}$ on the $x-y$ plane. We consider the coordinate transformation corresponding to the accelerated atom in the Rindler frame from Eq. \eqref{eq:CT-R-2}, with the spatial coordinates $\vec{x}_{\perp}$ remaining the same. The field mode solution, in this case, is given in Eq. \eqref{eq:Phi-3p1-Atom-Mirror}. With this field mode solution and the expression of the de-excitation probability \eqref{eq:TP-deex-gen} we obtain
\begin{eqnarray}\label{eq:DexProb-3p1-AccAtm-1}
    \mathcal{P}^{de}_{\nu}(\omega) 
    &=& \frac{g^2}{(2\pi)^3\,2\nu}\,\bigg|\int_{-\infty}^{\infty} d\tau\,\Big[\exp{\Big\{\frac{i\,k_{z}\,c^2}{a}\cosh{\Big(\frac{a\,\tau}{c}\Big)}-i\,k_{z}\,z_{0}\Big\}}~\exp{\Big\{\frac{i\,\nu\,c}{a}\sinh{\Big(\frac{a\,\tau}{c}\Big)}-i\,\omega\,\tau\Big\}}-c.c.\Big]\bigg|^2~.
\end{eqnarray}
With a change of variables $x=e^{a\,\tau/c}$, and then integrating over $x$ we obtain the de-excitation probability 
\begin{eqnarray}\label{eq:DexProb-3p1-AccAtm-2}
    \mathcal{P}^{de}_{\nu}(\omega) &=& \frac{c^2\,g^2}{(2\pi)^3\,2\,\nu\,a^2}\,\bigg|\pi\,e^{-i\,k_{z}\,z_{0}}\,\Big(\frac{\nu-c\,k_{z}}{\nu+c\,k_{z}}\Big)^{-\frac{i\,\omega\,c}{2\,a}}\,\mathbb{Y}\Big[\frac{i\,\omega\,c}{a},-\frac{c\sqrt{\nu^2-k_{z}^2\,c^2}}{a}\Big]-c.c.\bigg|^2~.
\end{eqnarray}
This expression can be simplified to express a Planckian distribution for large acceleration $a$. In this asymptotic scenario $\omega\,c/a\ll 1$, and we utilize the asymptotic expansion of the Bessel function $\mathbb{Y}_{\mu}(z)\sim -[(z/2)^{\mu}\,\cos{(\pi\,\mu)}\,\Gamma(-\mu)+(z/2)^{-\mu}\,\Gamma(\mu)]/\pi$ for $(z\ll 1)$. We obtain the de-excitation probability
\begin{eqnarray}\label{eq:DexProb-3p1-AccAtm-3}
    \mathcal{P}^{de}_{\nu}(\omega) &\simeq& \frac{c\,g^2}{2\,\pi^2\nu\,\omega\,a}~\frac{1}{1-e^{-2\pi\omega\,c/a}}~\Big|\sin{(k_{z}\,z_{0}-\chi_{1})}+\sin{(k_{z}\,z_{0}-\chi_{2})}\Big|^2~.
\end{eqnarray}
In the above expression we denote $\chi_{j}=\mathrm{Arg}[\varkappa_{j}]$ for $j$ being either $1$ or $2$, and $\varkappa_{1}=\Gamma(-i\,\omega\,c/a)\,\{c(\nu+c\,k_{z})/(2a)\}^{i\,\omega\,c/a}$ and $\varkappa_{2}=\Gamma(i\,\omega\,c/a)\,\{c(\nu-c\,k_{z})/(2a)\}^{-i\,\omega\,c/a}$. Considering large acceleration $a$ the expression of de-excitation probability can be further simplified to give
\begin{eqnarray}\label{eq:DexProb-3p1-AccAtm-4}
    \mathcal{P}^{de}_{\nu}(\omega) &\simeq& \frac{2\,c\,g^2}{\pi^2\nu\,\omega\,a}~\frac{\cos^2{(k_{z}\,z_{0})}}{1-e^{-2\pi\omega\,c/a}}~\sin^{2}{\left(\chi_{0}\right)}~.
\end{eqnarray}
In the above de-excitation probability $\chi_{0}\sim \mathcal{O}(1/a^2)$, and the entire expression can be obtained from the excitation probability of Eq. \eqref{eq:ExProb-3p1-AccAtm-5} by changing the sign of the atomic frequency, i.e., by $\omega\to -\omega$.

\subsubsection{When the mirror is in uniform acceleration}

Here we consider the atom static at $z=z_{0}$ and the mirror in uniform acceleration along the $+z$ direction. We also consider the coordinate transformation for the accelerated frame from Eq. \eqref{eq:CT-R}, and field mode solutions in the Rindler frame from Eq. \eqref{eq:Rind-FldMode-3p1}. With the asymptotic expansion of the Bessel function from Eqs. \eqref{eq:ExProb-3p1-AccMirr-BsslK-asymp1} and \eqref{eq:ExProb-3p1-AccMirr-BsslK-asymp2}, corresponding to large acceleration, the mode function gets simplified, which is already mentioned in eq. \eqref{eq:ExProb-3p1-AccMirr-Phi}. We utilize this expression of \eqref{eq:ExProb-3p1-AccMirr-Phi} along with the expression of the de-excitation probability from Eq. \eqref{eq:TP-deex-gen}, and the relation $\xi\pm c\,\tau = (c^2/a)\,\ln{\left[c^2(z\pm c\,t)/a\right]}$, to obtain
\begin{eqnarray}\label{eq:DexProb-3p1-AccMirr-1}
    \mathcal{P}^{de}_{\nu}(\omega) &=& \frac{g^2\,c\,\sinh{(\pi\,\nu\,c/a)}}{16\pi^4\,a}~\nonumber\\
    ~&\times&\bigg| \Gamma\Big(\frac{i\,\nu\,c}{a}\Big)\,\bigg(\frac{k_{\perp}}{2a/c^2}\bigg)^{-\frac{i\,\nu\,c}{a}}\, \int_{-\infty}^{z_{0}/c}dt\,\exp{\Big[-\frac{i\,\nu\,c}{a}\,\ln{\Big\{\frac{a}{c^2}(z-c\,t)\Big\}}-i\omega\,t\Big]} \nonumber\\
~&+& \Gamma\Big(-\frac{i\,\nu\,c}{a}\Big)\,\bigg(\frac{k_{\perp}}{2a/c^2}\bigg)^{\frac{i\,\nu\,c}{a}}\, \int_{-z_{0}/c}^{\infty}dt\, \exp{\Big[\frac{i\,\nu\,c}{a}\,\ln{\Big\{\frac{a}{c^2}(z+c\,t)\Big\}}-i\omega\,t\Big]}\nonumber\\
~&-& \Gamma\Big(\frac{i\,\nu\,c}{a}\Big)\,\bigg(\frac{k_{\perp}}{2a/c^2}\bigg)^{-\frac{i\,\nu\,c}{a}}\, \int_{-z_{0}/c}^{\infty}dt\, \exp{\Big[\frac{i\,\nu\,c}{a}\,\ln{\Big\{\frac{a}{c^2}(z+c\,t)\Big\}}-i\omega\,t\Big]} \nonumber\\
~&-& \Gamma\Big(-\frac{i\,\nu\,c}{a}\Big)\,\bigg(\frac{k_{\perp}}{2a/c^2}\bigg)^{\frac{i\,\nu\,c}{a}}\, \int_{-\infty}^{z_{0}/c}dt\,\exp{\Big[-\frac{i\,\nu\,c}{a}\,\ln{\Big\{\frac{a}{c^2}(z-c\,t)\Big\}}-i\omega\,t\Big]}\bigg|^2~.\label{eq:3p1-atomic-mirrorAcc-8b}
\end{eqnarray}
In the above expression, we consider a change of variables $x_{\pm}=\omega(t\pm z_{0}/c)$. Then with a further change of $x_{-}\to (-x_{-})$ the above expression is simplified to
\begin{eqnarray}\label{eq:DexProb-3p1-AccMirr-2}
    \mathcal{P}^{de}_{\nu}(\omega) &=& \frac{g^2\,c\,\sinh{(\pi\,\nu\,c/a)}}{4\pi^4\,a\,\omega^2}~\mathrm{Im}\bigg[\Gamma\Big(-\frac{i\,\nu\,c}{a}\Big)\,\bigg(\frac{k_{\perp}c^2}{2a}\bigg)^{\frac{i\,\nu\,c}{a}}\bigg]^2\,\nonumber\\
    ~&\times&\bigg| e^{-\frac{i\omega z_{0}}{c}} \int_{0}^{\infty}dx_{-}~\Big(\frac{a\,x_{-}}{\omega\,c}\Big)^{-\frac{i\nu\,c}{a}}\,e^{i\,x_{-}}-e^{\frac{i\omega z_{0}}{c}} \int_{0}^{\infty}dx_{+}~\Big(\frac{a\,x_{+}}{\omega\,c}\Big)^{\frac{i\nu\,c}{a}}\,e^{-i\,x_{+}}\bigg|^2 \nonumber\\
    ~&=&\frac{g^2\,c\,\sinh{(\pi\,\nu\,c/a)}}{4\pi^4\,a\,\omega^2}~\mathrm{Im}\bigg[\Gamma\Big(-\frac{i\,\nu\,c}{a}\Big)\,\bigg(\frac{k_{\perp}c^2}{2a}\bigg)^{\frac{i\,\nu\,c}{a}}\bigg]^2~
    \mathrm{Im}\bigg[e^{\frac{i\omega z_{0}}{c}}\,\Gamma\Big(\frac{i\,\nu\,c}{a}\Big)\,\bigg(\frac{a}{\omega\,c}\bigg)^{\frac{i\,\nu\,c}{a}}\bigg]^2\,\frac{\nu^2\,c^2\,e^{\pi\,\nu\,c/a}}{a^2}\nonumber\\
    ~&=& \frac{2\,g^2\,c\,\sin^2{(\Bar{\chi}_{1})}}{\pi^2\,a\,\omega^2}~\frac{\sin^2{(\Bar{\chi}_{2}+\omega\,z_{0}/c)}}{1-e^{-2\,\pi\,\nu\,c/a}}~.
\end{eqnarray}
In the above expression, the quantities $\Bar{\chi}_{j}$ are given by $\Bar{\chi}_{j}=\mathrm{Arg}[\Bar{\varkappa}_{j}]$, with $\Bar{\varkappa}_{1}=\Gamma(-i\,\nu\,c/a)\,\{k_{\perp} c^2/(2a)\}^{i\,\nu\,c/a}$ and $\Bar{\varkappa }_{2}= \Gamma(i\, \nu\,c/a)\,\{a/(\omega\,c)\}^{i\,\nu\,c/a}$. 
It is to be noted that one cannot obtain the above expression for the de-excitation probability from the expression of the atomic excitation of \eqref{eq:ExProb-3p1-AccMirr-3} by a sign change in $\nu$, i.e., by making $\nu\to -\nu$ in it, see Table \ref{tab:Obs2}.

\begin{table}[h!]
    \centering
    \begin{tabular}{ ||p{2.0cm}|p{4.0cm}|p{3.5cm}|p{6.7cm}||  }
 \hline
 \multicolumn{4}{|c|}{A tabular list of our observations from \ref{sec:AccAtom-E-WBound} and \ref{sec:AccAtom-DE-WBound}} \\
 \hline
 Dimensions & Excitation/De-excitation & Atom accelerated/static & Transition probabilities $\mathcal{P}^{ex/de}_{\nu}(\omega)$ \\
 \hline
 \multirow{2}{*}{$\mathbf{(1+1)}$} & Excitation & Accelerated & $\mathlarger{\mathlarger{\mathcal{P}^{ex}_{\nu}(\omega) = \frac{2\,c\,g^2}{a\,\omega\,\nu}\,\frac{\sin^2{(\nu\,z_{0}/c+\varphi_{1})}}{e^{2\pi\,\omega\,c/a}-1}}}$ \\\cline{3-4}
 &  & Static & $\mathlarger{\mathlarger{\mathcal{P}^{ex}_{\nu}(\omega) = \frac{2\,c\,g^2}{a\,\omega^2}\,\frac{\sin^2{(\omega\,z_{0}/c+\varphi_{2})}}{e^{2\pi\,\nu\,c/a}-1}}}$ \\\cline{2-4}
 \multirow{2}{*}{} & De-excitation & Accelerated & $\mathlarger{\mathlarger{\mathcal{P}^{de}_{\nu}(\omega)
    = \frac{2\,c\,g^2}{a\,\omega\,\nu}\,\frac{\sin^2{(\nu\,z_{0}/c+\Bar{\varphi}_{1})}}{1-e^{-2\pi\,c\,\omega/a}}}}$ \\\cline{3-4}
 &  & Static & $\mathlarger{\mathlarger{\mathcal{P}^{de}_{\nu}(\omega) = \frac{2\,c\,g^2}{\omega^2\,a}\,\frac{\sin^2{(\omega\,z_{0}/c+\Bar{\varphi}_{2})}}{1-e^{-2\pi\,c\,\nu/a}}}}$ \\
 \hline
 \multirow{2}{*}{$\mathbf{(3+1)}$} & Excitation & Accelerated & $\mathlarger{\mathlarger{\mathcal{P}^{ex}_{\nu}(\omega) \simeq \frac{2\,c\, g^{2}}{\pi^{2}\omega\,\nu\, a}~\frac{\cos^{2}{(k_{z}z_{0})}}{e^{2\pi\omega\,c/a}-1}\,\sin^{2}{\left(\psi_{0}\right)}}}$ \\\cline{3-4}
 &  & Static & $\mathlarger{\mathlarger{\mathcal{P}^{ex}_{\nu}(\omega) \simeq \frac{2\,g^2\,c}{\pi^2\,a\,\omega^2}\, \frac{\cos^2{\Big(\frac{\omega\,z_{0}}{c}\Big)}}{e^{\frac{2\pi\,\nu\,c}{a}}-1}}}$ \\\cline{2-4}
 \multirow{2}{*}{} & De-excitation & Accelerated &  $\mathlarger{\mathlarger{\mathcal{P}^{de}_{\nu}(\omega) \simeq \frac{2\,c\,g^2}{\pi^2\nu\,\omega\,a}~\frac{\cos^2{(k_{z}\,z_{0})}}{1-e^{-2\pi\omega\,c/a}}~\sin^{2}{\left(\chi_{0}\right)}}}$\\\cline{3-4}
 &  & Static & $\mathlarger{\mathlarger{\mathcal{P}^{de}_{\nu}(\omega) = \frac{2\,g^2\,c\,\sin^2{(\Bar{\chi}_{1})}}{\pi^2\,a\,\omega^2}~\frac{\sin^2{(\Bar{\chi}_{2}+\omega\,z_{0}/c)}}{1-e^{-2\,\pi\,\nu\,c/a}}}}$ \\
 \hline
\end{tabular}
    \caption{The above table lists our key observations regarding the atomic transitions in the presence of a mirror in $(1+1)$ and $(3+1)$ dimensions. We have listed both the atomic excitation and de-excitation probabilities from Secs. \ref{sec:AccAtom-E-WBound} and \ref{sec:AccAtom-DE-WBound}.}
    \label{tab:Obs2}
\end{table}

Our observations imply that in $(3+1)$ dimensions, in large acceleration limit, and when the atom is uniformly accelerated, the atomic excitation and de-excitation probabilities get reduced with increasing acceleration. For instance, one can see that (with the help of Eq. \eqref{eq:3p1-atomic-atomAcc-4a-3}) the atomic excitation \eqref{eq:3p1-atomic-atomAcc-4a-2} and de-excitation \eqref{eq:3p1-DeEx-atomAcc-2} probabilities without a boundary behave as $\mathcal{P}_{\nu}(\omega)\sim\mathcal{O}(1/a^2)$ for large accelerations. At the same time, in the presence of a boundary, the transition probabilities behave as $\mathcal{P}_{\nu}(\omega)\sim\mathcal{O}(1/a^4)$, see Eqs. \eqref{eq:ExProb-3p1-AccAtm-5} and \eqref{eq:DexProb-3p1-AccAtm-4}. On the contrary, when the atom is static the excitation and de-excitation probabilities tend to become independent of acceleration in the limit of large accelerations. For instance, from Eq. \eqref{eq:3p1-Ex-atomStatc-2} we observe that the atomic excitation probability, corresponding to a static atom without a boundary in $(3+1)$ dimensions, behaves as $\mathcal{P}^{ex}_{\nu}(\omega) \simeq \{g^{2}c/(4\pi^{3}\omega^{2}\nu)\}\,(1-\omega^{2}\,z_{0}^{2}/(2\,c^{2}))^{2}$, which is independent of the acceleration. Similarly in the presence of an accelerated mirror the excitation probability becomes $\mathcal{P}^{ex}_{\nu}(\omega) = \{g^2\,c/(\pi^3\,\omega^2\,\nu)\}\, \, \cos^2{(\omega \,z_{0}/c)}$ in the large acceleration limit. These observations remain the same with de-excitations. Notice that in all $(1+1)$ dimensional settings the transition probabilities become independent of the acceleration in a large acceleration limit. Here we would like to mention that we can never take $a\to\infty$ as the coordinate transformation \eqref{eq:CT-R-2} becomes invalid in this limit, and thus we can only talk about the scenarios when the acceleration is considerably large.

\section{Excitation to de-excitation ratio}\label{sec:Ex-Deex-ratio}

Usually, a system is found in an arbitrary quantum state. Therefore, initially, our two-level atom is more likely to be found in a superposition state, like $|\Psi_{in}\rangle= C_{0}\, |\omega_{0}\rangle + C_{1}\,|\omega_{1} \rangle$. Such an atom when interacts with scalar modes (see the interaction term in Eq. (\ref{eq:H-int})) which are in the vacuum state, then the atom will be again found in a linear superposition state, but with different superposition coefficients. Consequently, the probability of finding in excited and ground states will change, which can be interpreted as excitation and de-excitation probabilities. Hence for such a state, the excitation to de-excitation ratio (EDR) can be a relevant quantity to investigate. Such a quantity was introduced earlier in \cite{Garay:2016cpf} for a different purpose.

In this section, we would like to discuss this EDR in our current scenario, which we define as the ratio of $\mathcal{P}^{ex}_{\nu}(\omega)$ and $\mathcal{P}^{de}_{\nu}(\omega)$, to understand whether in terms of this quantity one can gain additional insight regarding the entire scenario. Our main aim here is to check whether in this ratio EDR we can get any equivalence between two specific scenarios one involving an accelerated atom and the other a static atom as the atomic and field frequency becomes the same. In $(1+1)$ dimensions this equivalence is observed even at the level of transition probabilities $\mathcal{P}^{ex}_{\nu}(\omega)$ and $\mathcal{P}^{de}_{\nu}(\omega)$. Therefore, here we shall focus only on the $(3+1)$ dimensional scenario. Here we define EDR as
\begin{eqnarray}\label{eq:EDR-Gen}
    \mathcal{R}_{\nu}(\omega) = \frac{\mathcal{P}^{ex}_{\nu}(\omega)}{\mathcal{P}^{de}_{\nu}(\omega)}~.
\end{eqnarray}
In $(3+1)$ dimensions without any boundary when the atom is accelerated, we take the expressions of $\mathcal{P}^{ex}_{\nu}(\omega)$ and $\mathcal{P}^{de}_{\nu}(\omega)$ from Eqs. \eqref{eq:3p1-atomic-atomAcc-4a-2} and \eqref{eq:3p1-DeEx-atomAcc-2}, and obtain the EDR as
\begin{eqnarray}\label{eq:EDR-AccsAtom}
    \mathcal{R}_{\nu}(\omega) = e^{-2\pi\omega\,c/a}~.
\end{eqnarray}
%
At the same time when the atom is static and responds to the Rindler vacuum, we take the expressions of $\mathcal{P}^{ex}_{\nu}(\omega)$ and $\mathcal{P}^{de}_{\nu}(\omega)$ from Eqs. \eqref{eq:3p1-Ex-atomStatc} and \eqref{eq:3p1-DeEx-atomStatc}, and obtain the EDR as
\begin{eqnarray}\label{eq:EDR-StAtm}
    \mathcal{R}_{\nu}(\omega) = e^{-2\pi\nu\,c/a}~.
\end{eqnarray}
It is to be noted that the above expressions of EDR in Eqs. \eqref{eq:EDR-AccsAtom} and \eqref{eq:EDR-StAtm} are obtained considering the excitation and de-excitation probabilities for large accelerations. From Eqs. \eqref{eq:EDR-AccsAtom} and \eqref{eq:EDR-StAtm} we observe that the EDR in $(3+1)$ dimensions for an accelerated atom (responding to the Minkowski vacuum) is the same as the EDR for a static atom (responding to the Rindler vacuum) when the atomic and field frequencies become the same, i.e., when we have $\omega=\nu$.

In $(3+1)$ dimensions when the atom is in uniform acceleration and the mirror is static we take the expressions of $\mathcal{P}^{ex}_{\nu}(\omega)$ and $\mathcal{P}^{de}_{\nu}(\omega)$ from Eqs. \eqref{eq:ExProb-3p1-AccAtm-5} and \eqref{eq:DexProb-3p1-AccAtm-4}, and get the expression of EDR as
\begin{eqnarray}\label{eq:EDR-AccsAtom-StMirr}
    \mathcal{R}_{\nu}(\omega) = e^{-2\pi\omega\,c/a}~.
\end{eqnarray}
At the same time, in $(3+1)$ dimensions when the atom is static and the mirror is in uniform acceleration, we take the expressions of $\mathcal{P}^{ex}_{\nu}(\omega)$ and $\mathcal{P}^{de}_{\nu}(\omega)$ from Eqs. \eqref{eq:ExProb-3p1-AccMirr-3} and \eqref{eq:DexProb-3p1-AccMirr-2}. In this scenario, the EDR is given by
\begin{eqnarray}\label{eq:EDR-AccsMirr-StAtm}
    \mathcal{R}_{\nu}(\omega) = e^{-2\pi\nu\,c/a}~.
\end{eqnarray}
In the above expressions of Eqs. \eqref{eq:EDR-AccsAtom-StMirr} and \eqref{eq:EDR-AccsMirr-StAtm} the EDR are obtained for excitation and de-excitation probabilities with large accelerations, see Appendix \ref{Appn:Asymp-Exp}. From Eqs. \eqref{eq:EDR-AccsAtom-StMirr} and \eqref{eq:EDR-AccsMirr-StAtm} one can observe that the EDR in $(3+1)$ dimensions for an accelerated atom with static mirror is the same as the EDR for a static atom with accelerated mirror when the atomic and field frequencies become the same, i.e., in the specific situation of $\omega=\nu$. We should note that this was not the case at the level of atomic excitations (see Eqs. \eqref{eq:ExProb-3p1-AccAtm-5} and \eqref{eq:ExProb-3p1-AccMirr-3}) or de-excitations (see Eqs. \eqref{eq:DexProb-3p1-AccAtm-4} and \eqref{eq:DexProb-3p1-AccMirr-2}).

\section{Discussion}\label{sec:discussion}

It is well-known that the thermal spectrum of particles observed in the Minkowski vacuum by a uniformly accelerated observer is equivalent to those in the Rindler vacuum as observed by a static detector, see \cite{Takagi:1986kn}. In this direction, in a similar set-up it is also seen that the excitation probability of a uniformly accelerated atom in the presence of a static mirror is identical to the excitation probability of a static atom in the presence of an accelerated mirror when the atomic and field frequencies are the same \cite{Svidzinsky:2018jkp}. Although obtained in $(1+1)$ dimensions this result enables one to presume that in $(3+1)$ dimensions too some of the features of the atomic spectra should be similar in equivalent scenarios. However, up to what level this equivalence remains true needs investigation. In this work, we have particularly tried to answer these questions. With this objective, we have considered atoms with and without the presence of reflecting boundaries and extensively studied their excitation in $(3+1)$ dimensions as the results in $(3+1)$ dimensions were not known. We have also recalled the $(1+1)$ dimensional results where relevant. Furthermore, in a first of its kind, we have investigated the atomic de-excitation probabilities in both $(1+1)$ and $(3+1)$ dimensions for the previously mentioned atomic set-ups, which corroborates many important results from the excitation probability regarding $(3+1)$ dimensional scenario and imparts significant insights.

The observations from previous literary works \cite{Svidzinsky:2018jkp} and our present work suggest that in $(1+1)$ dimensions the atomic excitation probabilities, corresponding to the accelerated and static atom scenarios, are the same when the atomic $(\omega)$ and field $(\nu)$ frequencies are equal, i.e., when $\omega=\nu$. In this regard, one can recall the expressions of \eqref{eq:Pex-Atom-Acc} and \eqref{eq:Pex-Atom-Stat} for atomic excitations without a boundary. At the same time, when the presence of a mirror is considered one can look into the expressions of atomic excitations from Eqs. \eqref{eq:ExProb-1p1-AccAtm} and \eqref{eq:ExProb-1p1-AccMirr}. In $(3+1)$ dimensions in general we could not find this equivalence even for atoms without a boundary. For instance, the atomic excitations in \eqref{eq:3p1-atomic-atomAcc-4a-2} and \eqref{eq:3p1-Ex-atomStatc}, which respectively correspond to the response of an accelerated atom in Minkowski vacuum and a static atom in Rindler vacuum, are not the same when $\omega=\nu$. Similarly, when there is a boundary present this equivalence is not present. For instance, the expressions \eqref{eq:ExProb-3p1-AccAtm-5} and \eqref{eq:ExProb-3p1-AccMirr-5}, which respectively denote the excitations of accelerated and static atoms in the presence of static and accelerated mirrors, are not the same for $\omega=\nu$. However, we have seen that even in these $(3+1)$ dimensional excitation probabilities, the thermal factor part remains equivalent under the condition $\omega=\nu$ between different scenarios (corresponding to one Rindler and another Rindler equivalent set-up). We inferred that the difference in these excitation probabilities is due to the multiplicative factors (periodic in the field or atomic frequencies) of the thermal factor. When the atom is uniformly accelerated, this multiplicative factor depends on the field frequency $\nu$ with the information of the direction along which the atom is accelerated, see Eqs. \eqref{eq:3p1-atomic-atomAcc-4a-2} and \eqref{eq:ExProb-3p1-AccAtm-5}. In contrast, when the atom is static this factor depends on the atomic frequency $\omega$ without any directional information, see Eqs. \eqref{eq:3p1-Ex-atomStatc}  and \eqref{eq:ExProb-3p1-AccMirr-4}. In the first set of scenarios, this directional dependence of the excitation probabilities debars one to equate these excitation probabilities with the second scenario when $\nu=\omega$.

We have also studied the de-excitation probability corresponding to each of the concerned scenarios. Our observations confirm that in $(1+1)$ dimensions and without a boundary, the de-excitation in an accelerated atom is identical to the de-excitation of a static atom responding to the Rindler vacuum when $\omega=\nu$, see Eqs. \eqref{eq:Pdex-Atom-Acc} and \eqref{eq:Pdex-Atom-Stat}. At the same time, in the presence of a mirror in $(1+1)$ dimensions, the de-excitations for the accelerated and static atoms are the same when $\omega=\nu$, see Eqs. \eqref{eq:DexProb-1p1-AccAtm} and \eqref{eq:DexProb-1p1-AccMirr-2}. In $(3+1)$ dimensions no such general equivalence in the de-excitation probabilities is observed between different scenarios of atomic motion when $\omega=\nu$. In this regard, one can compare Eqs. \eqref{eq:3p1-DeEx-atomAcc-2} and \eqref{eq:3p1-DeEx-atomStatc} when there is no boundary, and \eqref{eq:DexProb-3p1-AccAtm-4} and \eqref{eq:DexProb-3p1-AccMirr-2} when there is a mirror. In $(3+1)$ dimensions like the excitation probability here also one can notice that the thermal factors become identical when $\omega=\nu$. However, the multiplicative factor to the thermal part has a directional dependence for the cases when the atoms are accelerated, which hinders its equivalence to the cases with static atoms.

Our other observations imply in both $(1+1)$ and $(3+1)$ spacetime dimensions and with and without boundary if the atom is in uniform acceleration the de-excitation probability can be obtained from the excitation probability by changing the sign of the atomic frequency, i.e., by making $\omega\to -\omega$. In this regard, see and compare \eqref{eq:Pex-Atom-Acc} and \eqref{eq:Pdex-Atom-Acc} for $(1+1)$ dimensions, and \eqref{eq:3p1-atomic-atomAcc-4a-2} and \eqref{eq:3p1-DeEx-atomAcc-2} for $(3+1)$ dimensions when there is no boundary. Similarly, in the presence of a boundary compare \eqref{eq:ExProb-1p1-AccAtm} and \eqref{eq:DexProb-1p1-AccAtm} for $(1+1)$ dimensions, and \eqref{eq:ExProb-3p1-AccAtm-5} and \eqref{eq:DexProb-3p1-AccAtm-4} for $(3+1)$ dimensions. At the same time, when the atom is static, and it responds to the Rindler vacuum, or it is in the presence of an accelerated mirror, one cannot obtain the atomic de-excitation probability from the excitation probability by changing the sign of the atomic or field frequencies in all scenarios, i.e., by making $\omega\to -\omega$ or $\nu\to -\nu$. For instance, in $(1+1)$ dimensions when the atom is static and the mirror is in uniform acceleration, we get the de-excitation probability by making $\nu\to -\nu$ in the excitation probability. In contrast, by making a similar sign change, we cannot get the de-excitation from the excitation probability when the atom is static in $(1+1)$ and $(3+1)$ dimensions without a boundary or in the presence of a mirror in $(3+1)$ dimensions.

It is evident that when the atom is accelerated the physics is more straightforward, as one can get the de-excitation probability from excitation just by changing the sign of the atomic frequency $\omega$. This is not the case for a static atom in an accelerated ambiance. However, this observation can be treated to indicate a deeper equivalence as in the second scenario the field frequency $\nu$ can sometimes play a similar role to the atomic frequency $\omega$ from the first scenario. The $(3+1)$ dimensional scenario also opens up an avenue to, in principle, verify these predictions in an experimental setting. In this regard, one can take help from several proposals concerning the realization of an atomic probe  \cite{Aspachs:2010hh, Rodriguez-Laguna:2016kri, Gooding:2020scc, Barman:2024jpc}. 
We have also seen that though the entire transition probabilities from different scenarios are not equivalent in $(3+1)$ dimensions, the excitation to de-excitation ratio, the EDR, is the same when $\omega=\nu$ for large accelerations. In this regard, see \eqref{eq:EDR-AccsAtom} and \eqref{eq:EDR-StAtm} for atoms without a boundary, and \eqref{eq:EDR-AccsAtom-StMirr} and \eqref{eq:EDR-AccsMirr-StAtm} in the presence of boundary. Now in a realistic scenario, the $(3+1)$ dimensional results are likely to be observed in an experimental set-up. So we can conclude that when the equivalence of a uniformly accelerated atom to an accelerated environment is concerned it is more suitable to study the EDR than the individual excitation or de-excitation probabilities.
With similar objectives, an interesting investigation could be to understand how entangled quantum probes get affected when kept in accelerated motion as compared to the accelerated environment. In this regard, the works \cite{Reznik:2002fz, Koga:2018the, Barman:2021bbw, Barman:2021kwg, Barman:2022xht, Chowdhury:2021ieg, Barman:2022loh, Barman:2023rhd, Barman:2023aqk, K:2023oon} dealing with entanglement induced between quantum probes, and \cite{Arias:2015moa, Menezes:2015iva, Barman:2021oum, Barman:2022utm, Barman:2024vah, Arya:2024qke} involving the radiative process of entangled probes could be specifically useful.

The lack of equivalence between the atomic spectra for an accelerated atom as compared to a static atom in an accelerated ambience in $(3+1)$ dimensions can be physically interpreted in the following manner. When the atom is accelerated it sees the Minkowski vacuum to be populated by a thermal distribution of particles. As an accelerated observer sees these particles we call them the Rindler particles. When an atom is accelerated these Rindler particles, similarly the field as expressed with respect to the Minkowski vacuum, see the atom as it is, i.e., with acceleration along a particular direction. For an accelerated atom, the Minkowski modes in terms of the accelerated coordinates will contain the directional information even in a large acceleration limit. This is why in the excitation or de-excitation probability the information about this direction is encoded. Similarly, when a static atom sees the Rindler vacuum, we call the particles to be the Minkowski particles as the observer is now static. In this scenario, the field defined with respect to the Rindler vacuum loses this directional information for large acceleration which is evident from Eqs. \eqref{eq:Rind-FldMode-3p1-2}, \eqref{eq:ExProb-3p1-AccMirr-BsslK-asymp1}, and \eqref{eq:ExProb-3p1-AccMirr-BsslK-asymp2}. Therefore, the excitation or de-excitation of a static atom by absorbing or emitting a virtual Minkowski particle also does not concern this direction. We infer that this is the reason why in $(3+1)$ dimensions the above two scenarios are different. With this, we believe our work has provided significant insight regarding the equivalence between atomic spectra for accelerated atoms as compared to static atoms with respect to an accelerated ambience.

\begin{acknowledgments}

S.B. would like to thank the Science and Engineering Research Board (SERB), Government of India (GoI), for supporting this work through the National Post Doctoral Fellowship (N-PDF, File number: PDF/2022/000428). 

\end{acknowledgments}

\appendix
\section{Asymptotic and simplified forms of different expressions}\label{Appn:Asymp-Exp}
In this section of the appendix, we present some simplified asymptotic forms of the expressions for transition probabilities, which can ease the comparison of excitation and de-excitation probabilities and help one to evaluate the EDR. In this regard, we shall concern ourselves with the scenario with a mirror, as the without-mirror scenario is straightforward to compare. For instance, one can check Table \ref{tab:Obs1}, and observe that $\cos^{2}{\varphi}_{ex} =\cos^{2}{\varphi}_{de}$ and $\cos^{2}{\Bar{\varphi}_{ex}} =\cos^{2}{\Bar{\varphi}_{de}}$ from the expressions of $\varphi_{ex/de}$ and $\Bar{\varphi}_{ex/de}$ respectively. Then the EDR corresponding to this scenario without a mirror becomes trivial to calculate. At the same time, one can notice that the issues related to the non-equivalence of excitation probabilities in the presence of a mirror arise only in $(3+1)$ dimensions, for which we have evaluated the EDR. Therefore, let us focus on this scenario of $(3+1)$ dimensions in the presence of a mirror.\vspace{0.2cm}

The asymptotic expression of $\psi_{0}$ in $\mathcal{P}^{ex}_{\nu}(\omega)$ from Eq. \eqref{eq:ExProb-3p1-AccAtm-5} and Table \ref{tab:Obs2} is given by $\psi_{0}\simeq (c\,\omega\,\rho/2a)^2$. Similarly, one can show that the asymptotic expression of $\chi_{0}$ in $\mathcal{P}^{de}_{\nu}(\omega)$ from Eq. \eqref{eq:DexProb-3p1-AccAtm-4} and Table \ref{tab:Obs2} is also given by $\chi_{0}\simeq (c\,\omega\,\rho/2a)^2$, where $\rho$ is obtained from
\begin{eqnarray}
    \rho^2 &=& \log \left(\frac{\nu + c\, k_{z}}{\nu -c\, k_{z}}\right) \left[\gamma + \log \left(\frac{c\,\sqrt{\nu ^2-c\,^2 k_{z}^2}}{2\, a}\right) \right]~.
\end{eqnarray}
Therefore, it is easy to understand that when the atom is in uniform acceleration and the mirror is static the EDR will be given by Eq. \eqref{eq:EDR-AccsAtom-StMirr}. On the other hand, let us series expand the de-excitation probability $\mathcal{P}^{de}_{\nu}(\omega)$ from Eq. \eqref{eq:DexProb-3p1-AccMirr-2} for large $a$ and check whether it has some resemblance with the asymptotic expression of $\mathcal{P}^{ex}_{\nu}(\omega)$ from Eq. \eqref{eq:ExProb-3p1-AccMirr-4}. In particular, from Eq. \eqref{eq:DexProb-3p1-AccMirr-2} and for large $a$, one can obtain $\Bar{\varkappa}_{1}\simeq (i\,a/c\,\nu)$ and $\Bar{\varkappa}_{2}\simeq -(i\,a/c\,\nu)$. Then for large $a$ we will have the asymptotic forms $\Bar{\chi}_{1}\simeq \pi/2$ and $\Bar{\chi}_{2}\simeq -\pi/2$, and the de-excitation probability of Eq. \eqref{eq:DexProb-3p1-AccMirr-2} will become
\begin{eqnarray}
    \mathcal{P}^{de}_{\nu}(\omega) &\simeq& \frac{2\,g^2\,c}{\pi^2\,a\,\omega^2}~\frac{\cos^2{(\omega\,z_{0}/c)}}{1-e^{-2\,\pi\,\nu\,c/a}}~.
\end{eqnarray}
If we utilize this expression with the expression of $\mathcal{P}^{ex}_{\nu}(\omega)$ from Table \ref{tab:Obs2} we shall get the EDR of Eq. \eqref{eq:EDR-AccsMirr-StAtm}.

\bibliographystyle{utphys1.bst}

\bibliography{bibtexfile}

\end{document}